\documentclass[
    a4paper,notitlepage,
    twocolumn,
    nofootinbib,
    amsmath,amssymb
]{revtex4-2}

\usepackage{graphicx}
\graphicspath{{figures/}}

\usepackage[usenames,dvipsnames]{xcolor} 

\usepackage[
    colorlinks = true,
    urlcolor = MidnightBlue,
    linkcolor = Maroon,
    citecolor = Maroon
]{hyperref}

\hypersetup{
    pdfauthor = {A. V. Glushkov, L. T. Ksenofontov, K. G. Lebedev, A. Saburov},
    pdftitle = {On the calibration of ultra-high energy EASs at the Yakutsk array and Telescope Array},
    pdfsubject = {ultra-high energy cosmic rays, extensive air showers},
    pdfkeywords = {cosmic ray spectrum, energy estimation, SD response, Yakutsk array, Telescope Array}
}

\usepackage{lineno}

\usepackage{etoolbox}
\makeatletter
\patchcmd{\frontmatter@RRAP@format}{(}{}{}{}
\patchcmd{\frontmatter@RRAP@format}{)}{}{}{}
\makeatother

\newcommand{\eqn}[1]{(\ref{#1})}
\newcommand{\fign}[1]{Fig.~\ref{#1}}
\newcommand{\pancap}[1]{\textsl{(#1)}}            
\newcommand{\panref}[2]{Fig.~\ref{#1}\textsl{#2}} 

\newcommand{\nthick}{\thickmuskip=0mu} 
\def\thesection       {\arabic{section}}
\def\p@section        {}
\def\thesubsection    {\thesection.\arabic{subsection}}
\def\p@subsection     {\thesection.}

\def\p@subsubsection  {\thesection\,\thesubsection\,}

\makeatletter
\renewcommand \thefigure{\@arabic\c@figure}
\renewcommand \thetable{\@arabic\c@table}

\usepackage{array, multirow}

\newcolumntype{C}[1]{>{\centering\arraybackslash}p{#1}}   
\newcolumntype{R}[1]{>{\raggedleft\arraybackslash}p{#1}}  
\newcolumntype{L}[1]{>{\raggedright\arraybackslash}p{#1}} 

\newcommand{\E}{E_0}
\newcommand{\EFD}{E_{\text{FD}}}
\newcommand{\EFDOne}{E^1_{\text{FD}}}

\newcommand{\ESD}{E_{\text{SD}}}
\newcommand{\ESDOne}{E^1_{\text{SD}}}
\newcommand{\ESDSix}{E_{\text{SD}, 6}}
\newcommand{\ESDTen}{E_{\text{SD}, 10}}
\newcommand{\ESDCor}{E_{\text{CORSIKA, SD}}}

\newcommand{\vem}{\epsilon_1}       
\newcommand{\VEM}{\text{VEM}}

\newcommand{\submin}{_{\text{min}}}
\newcommand{\subscr}{_{\text{scr}}}
\newcommand{\submu}{_{\mu}}
\newcommand{\subgam}{_{\gamma}}

\newcommand{\ovrscr}{^{\text{scr}}}
\newcommand{\ovrsct}{^{\text{sct}}}
\newcommand{\subpair}{_{e\pm}}
\newcommand{\subcomp}{_{\delta e}}

\newcommand{\MuThr}{\varepsilon_{\mu, \text{thr.}}}
\newcommand{\ElThr}{\varepsilon_{e, \text{thr.}}}

\newcommand{\subthr}{_{\text{thr.}}}
\newcommand{\subch}{_{\text{ch}}}
\newcommand{\ovrRelUn}{^{\text{arb.u.}}}

\newcommand{\fcic}{f_{\text{CIC}}}

\newcommand{\xmax}{x_{\text{max}}}
\newcommand{\xobs}{x_{\text{obs.}}}
\newcommand{\Nmax}{N_{\text{max}}}

\newcommand{\sqrm}{m$^2$}     
\newcommand{\cubcm}{cm$^3$}   
\newcommand{\depth}{g/cm$^2$} 
\newcommand{\dens}{g/cm$^3$}  
\newcommand{\pdens}{m$^{-2}$} 
\newcommand{\degr}{^{\circ}}  
\newcommand{\edeposit}{\left[\frac{\text{MeV}}{\text{g/cm}^2}\right]} 

\newcommand{\qgsii}{{\sc qgsj}et-{\sc ii}.04}

\newcommand{\corsika}{{\sc corsika}}

\begin{document}

\title{On the calibration of ultra-high energy EASs at the Yakutsk array and Telescope Array}

\author{\bf A.\,V.\,Glushkov\textsuperscript{1)}}
\email{glushkov@ikfia.ysn.ru}

\author{A.\,V.\,Saburov}
\email{vs.tema@gmail.com}

\author{L.\,T.\,Ksenofontov\textsuperscript{1)}}
\email{ksenofon@ikfia.ysn.ru}

\author{K.\,G.\,Lebedev\textsuperscript{1)}}
\email{LebedevKG@ikfia.ysn.ru}

\affiliation{\normalfont{\textsuperscript{1)}Yu.\,G.\,Shafer Institute of Cosmophysical Reserach and Aeronomy of Siberian branch of the Russian Academy of Sciences, 31 Lenin ave., Yakutsk, 677027, Russia}}

\begin{abstract}
    \noindent
    {\bf Abstract~---} An analysis of calibrations of extensive air showers with zenith angles $\theta \le 50\degr$ and energies $\ESD \ge 10^{18.5}$~eV was carried out in experiments at the Yakutsk array and Telescope Array. The values of $\ESD$ were determined from particle densities measured with ground-based scintillation detectors at a distance $r = 800$~m from shower axis. Measured densities were compared with the values obtained in simulations preformed with the use of \corsika{} code within the framework of \qgsii{} hadron interaction model for primary protons. For showers with $\theta = 0\degr$ the $\ESD$ estimates of both arrays are very close, and their energy spectra are similar in shape and absolute value. Another Telescope Array calibration, based on measuring the EAS fluorescent radiation with optical detectors, gives underestimated energy $\EFD = 0.787 \times \ESD$.
\end{abstract}

\maketitle

\section{Introduction}

Ultra-high energy cosmic rays~--- cosmic rays (CR) with energy above $\nthick\sim 10^{14}$~eV~--- have been studied for more than 50 years. Much is already known about them (see review~\cite{b:1} for example). But since CR with these energies can only be studied by registering extensive air showers (EAS), cascades of secondary particles initiated by them in Earth's atmosphere, there is a great diversity in estimations of the energy of primary particles. It touches upon such important aspects as CR energy spectrum at energy $\E \ge 10^{17}$~eV~\cite{b:2, b:4} and the so-called ``muon puzzle'': a disagreement between theoretically predicted and experimentally measured muon flux densities in EAS, together with discrepancies between different experiments~\cite{b:5, b:6}.

\begin{figure}[!htb]
    \includegraphics[width=0.480\textwidth]{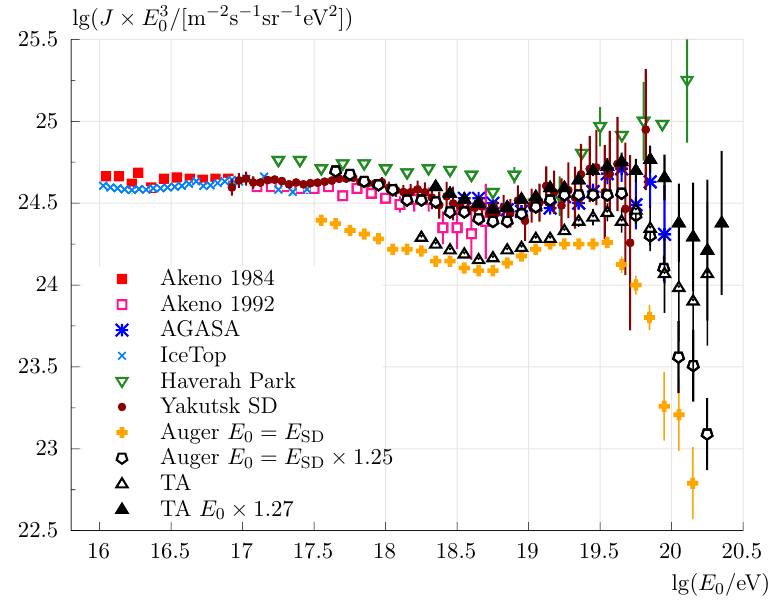}
    \caption{Differential energy spectrum of CR according to the data of different world arrays: the Yakutsk Array~\cite{b:2}, Akeno~\cite{b:7, b:8}, AGASA~\cite{b:9}, IceTop~\cite{b:10}, Auger~\cite{b:11}, TA~\cite{b:12} and Haverah Park~\cite{b:13}.}
    \label{f:1}
\end{figure}

Review~\cite{b:1} considers only The Pierre Auger Observatory (Auger) and Telescope Array (TA) experiments which measure CR energy using the same technique~--- from readings of optical detectors registering the fluorescent light emission accompanying the EAS development (fluorescent detectors~--- FD). The results of earlier experiments~--- The Yakutsk array (continuously operating since 1973 to this day), Havera Park, AGASA, etc.~--- are not mentioned. CR spectra obtained at these arrays~--- Akeno (1984, 1992)~\cite{b:7, b:8}, AGASA~\cite{b:9}, Ice Top~\cite{b:10}, Auger~\cite{b:11}, TA~\cite{b:12} and Haverah Park~\cite{b:13}~--- are shown in \fign{f:1}. It is seen that spectra of Auger and TA are roughly consistent with each other but contradict all other experiments. It was shown earlier~\cite{b:4, b:5} that CR energy measured by Auger~\cite{b:11} was underestimated by 25\%. It is also seen from the figure that the Auger spectrum with energy rescaling introduced in~\cite{b:5} does not contradict other experiments which estimate CR energy with alternative method~--- by measuring the lateral distribution function (LDF) of shower particles with surface detectors (SD) at observation level.

Work~\cite{b:14} provides a detailed description of methods for estimating primary energy and calculating energy spectra~\cite{b:12} presented in~\fign{f:1}. Both spectra were obtained from the readings of scintillation SDs located at the nodes of a rectangular grid with a side 1200~m. They measure densities of all shower particles at the observation level and are used to reconstruct the main EAS parameters: arrival direction, axis coordinates and primary energy $\E$. Calculations have revealed (see Section~4) that the $S(800)$ parameter, detector signals recorded at axis distance $r = 800$~m in showers with zenith angles $\theta = 0\degr$, is unambiguously connected to the primary energy $\ESD$ with the following relation:

\begin{equation}
    \ESD = \ESDOne \times S(800, 0\degr)^{1.025 \pm 0.010}~\text{[eV],}
    \label{eq:1}
\end{equation}
where $\ESDOne = (2.29 \pm 0.08) \times 10^{17}$~eV. The spectrum thus obtained is shown in \fign{f:1} with dark upward triangles. It does not contradict other data. Another spectrum, shown with upward open triangles, was obtained by estimating primary energy $\EFDOne$ of the same showers from readings of optical FDs. Values $\ESDOne$ and $\EFDOne$ were calculated using the Monte-Carlo method. They differ from each other by factor $\nthick\ESD/\EFD = 1.27$. Below we will consider this issue in more detail.

\section{Particle density measurement}

\subsection{Scintillation detectors}

\begin{figure}[!htb]
    \centering
    \includegraphics[width=0.480\textwidth]{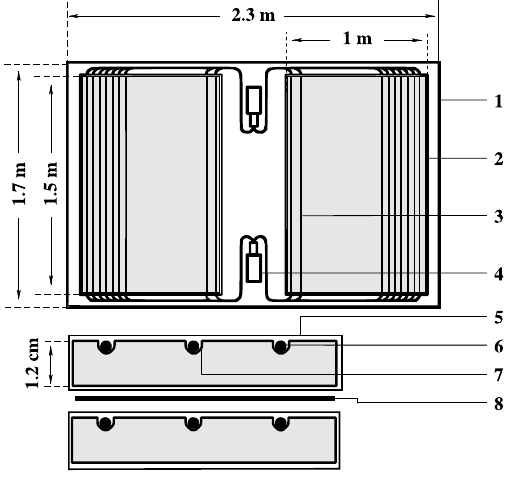}
    \caption{The TA SD (Fig.~2.8 in~\cite{b:14}): 1~--- container made of 1.2~mm thick steel sheet with dimensions $2.3 \times 1.7 \times 0.1$~m in which two 1.2~cm thick scintillation layers are stacked (2). Container~1 is placed under a 1.2-mm steel cover; 3~--- optical fibers transferring ionization flashes into PMT (4); 6~--- cross-section of an optical fiber (1~mm diameter); 7~--- slots for optical fibers (1.2~mm diameter). Both scintillation layers are wrapped in Tyvek (5); 8~--- separation plate made of 1~mm thick stainless steel. Density of scintillator is 1.032~\dens.}
    \label{f:2}
\end{figure}

The schematics of a standard scintillation SD constituting the ground array of TA is shown in \fign{f:2}. Polyvinyl toluene (C$_9$H$_{10}$) was chosen as scintillator plastic. In total there are 507 such detectors included in the events selection system. Two identical scintillation layers each with effective area $s = 1.5 \times 1 \times 2 = 3$~\sqrm{} are separated with a thin steel plate (8). They duplicate each other's function to monitor the reliability of EAS particle density measurements. Optical fibers (3) collect light produced in scintillators and shift its spectrum into the working range of a photomultiplier tube (PMT). Each scintillation layer is observed by a separate PMT.

\begin{figure}[!htb]
    \centering
    \includegraphics[width=0.480\textwidth]{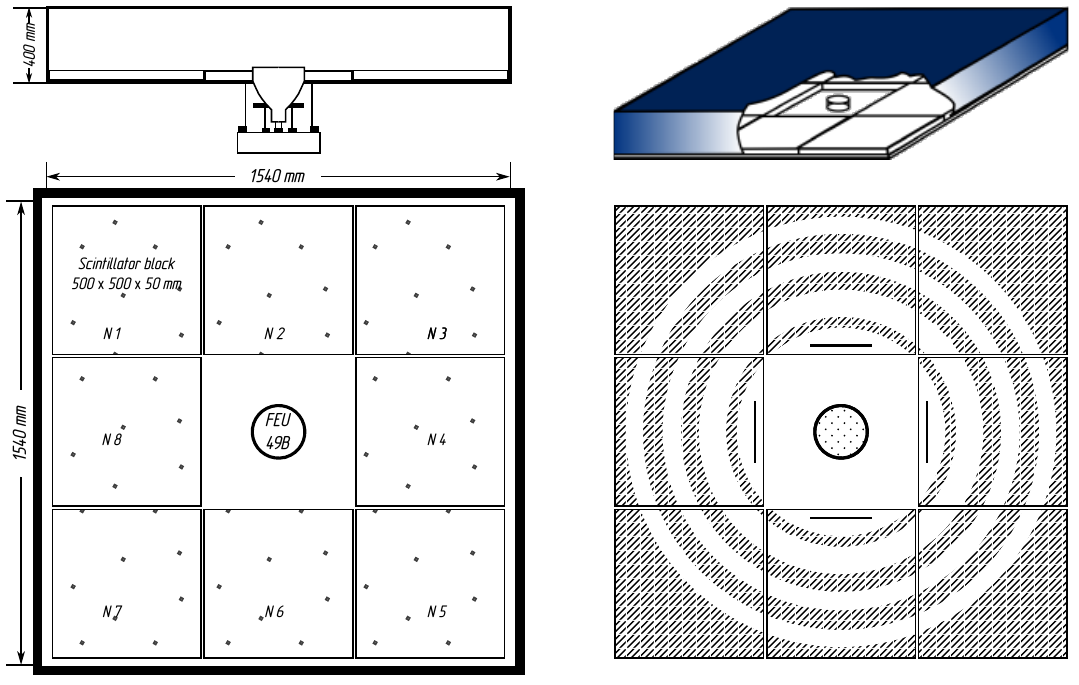}
    \caption{Standard scintillation detector of Yakutsk array with area $s = 0.25 \times 8 = 2$~\sqrm. Density of scintillator is 1.06~\dens. In the center PMT FEU-49 is mounted.}
    \label{f:3}
\end{figure}

For measurement of EAS particle flux the Yakutsk array utilizes 2-\sqrm{} detectors based on plastic scintillators made of polystyrene (C$_8$H$_8$) with luminescent additives ($\nthick\sim 2$\% p-terphenyl and $\nthick\sim 0.02$\% POPO) in the form of $50 \times 50 \times 5$-\cubcm{} blocks~\cite{b:5, b:6}. Eight such blocks are placed around the perimeter of the platform of a light-proof container as shown in \fign{f:3}. In the center of the platform a PMT FEU-49 is mounted. The light generated inside scintillator blocks enters the FEU-49 via diffusive reflection from inner surfaces of the cover which are coated with a special white paint. The cover of the container is made of 1.5-mm aluminium sheet. The maximum light yield is approximately at $440$~nm which fits well within spectral characteristics of FEU-49 and reflective properties of painted surfaces of the container. The glow duration of scintillator is about several nanoseconds. To increase the light collection, sides and bottom of each scintillator block were coated with white paint.

\subsection{Measurement units}

\paragraph*{Telescope Array.}

\begin{figure}[!htb]
    \centering
    \includegraphics[width=0.238\textwidth]{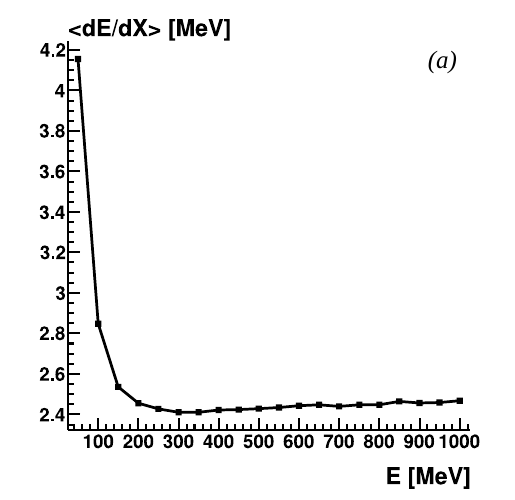}
    \includegraphics[width=0.238\textwidth]{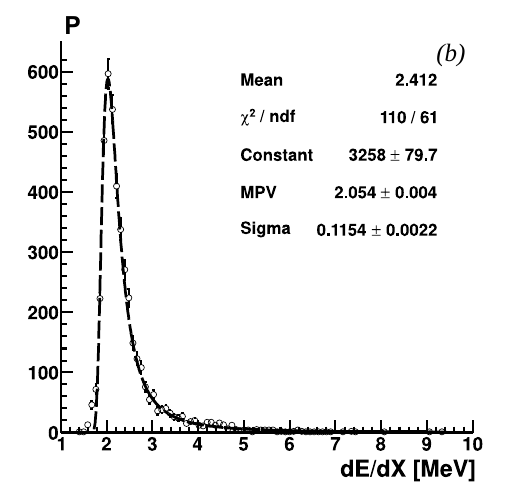}
    \caption{\pancap{a} The mean energy deposit of a vertical muon in a 1.2~cm thick layer of scintillator in the TA SD setup (Fig.~2.9 from~\cite{b:14}). \pancap{b} Energy losses of muons relative to the 300~MeV value (minimum ionization in \panref{f:4}{a}) in a 1.2~cm thick scintillator obtained via simulation with Geant4. MPV is for the most probable value. Dashed curve represents the Landau approximation~\cite{b:16} (Fig.~2.10 from~\cite{b:14}).}
    \label{f:4}
\end{figure}

In \panref{f:4}{a} the mean energy deposit is shown for a vertical muon traversing a 1.2-cm layer of scintillator (see~\fign{f:2}), obtained by simulating the SD response using Geant4 toolkit~\cite{b:15}. Near 300~MeV a wide minimum of ionization is observed. Histogram in \panref{f:4}{b} is approximated with the Landau distribution~\cite{b:16}. A small tail on the left arises from marginal effects. As a unit of response from a single vertical muon (Vertical Equivalent Muon~--- VEM) the following value is selected:

\begin{equation}
    \VEM = \vem(0\degr) = 2.05~\text{MeV,}
    \label{eq:2}
\end{equation}
which is the most probable energy deposit for a vertical muon with minimum ionization energy (300~MeV).

\paragraph*{The Yakutsk array.}

The particle density of EAS at Yakutsk array is measured in units of energy deposited by vertical relativistic muons in a 5~cm thick plastic scintillator with 1.06~\dens{} density~\cite{b:4, b:5}:

\begin{equation}
    \epsilon_1(0\degr) = 11.75~\text{MeV.}
    \label{eq:3}
\end{equation}

\begin{figure}[!htb]
    \centering
    \includegraphics[width=0.480\textwidth]{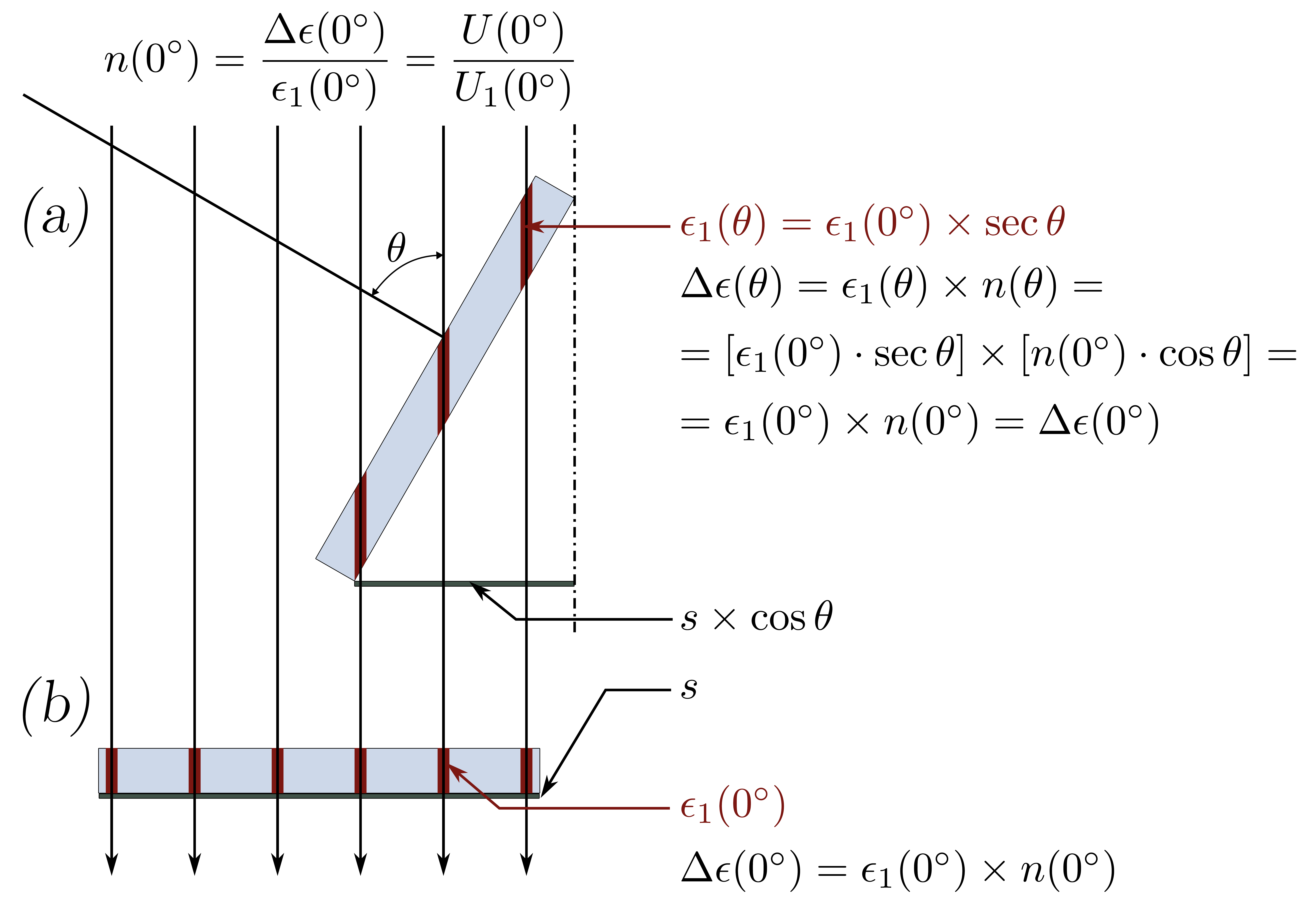}
    \caption{The diagram of the formation of total response $\Delta\epsilon$ during the passage of $n$ EAS particles arriving at different zenith angles ($\theta$) through a scintillation detector of the area $s$.}
    \label{f:5}
\end{figure}

Energy deposit \eqn{eq:3} is shown in \fign{f:5} with darkened tracks inside the scintillator block. It is spent by muon on ionization of the medium of scintillator and is re-emitted as a light flash. This flash is subsequently converted by PMT into electric pulse with the amplitude $U_1(0\degr)$~--- the level of a single particle (calibration level). The value $U_1(0\degr)$ is regularly measured by collecting the amplitude spectra from the scintillation detector. In \fign{f:5} the total energy deposit is shown in inclined $\Delta\epsilon(\theta)$ \textsl{(a)} and vertical $\Delta\epsilon(0\degr)$ \textsl{(b)} showers respectively. These deposits are the same at any zenith angles. The number of particles at a distance $r$ from the axis is determined with the formula:

\begin{equation}
    n(r, 0\degr) = \frac{
        \Delta\epsilon(r, 0\degr)
    }{
        \epsilon_1(0\degr)
    } = \frac{
        U(r, 0\degr)
    }{
        U_1(0\degr)
    }\text{.}
    \label{eq:4}
\end{equation}

The particle density of a shower with zenith angle $\theta$ that crossed the detector with surface area $s$ at axis distance $r$ subsequently equals to:

\begin{align}
    \rho(r, \theta) &= \frac{n(r, \theta)}{s(\theta)} = 
    \frac{n(r, 0\degr) \cdot \cos\theta}{s \cdot \cos\theta} = \nonumber \\
    &\quad = \frac{U(r)}{U_1(0\degr) \cdot s} = 
    \frac{n(r)}{s}~\text{[\pdens].}
    \label{eq:5}
\end{align}

It does not depend on the shower arrival angle since the amplitude on PMT's output doesn't change~\cite{b:4, b:5}.

\section{Simulation}

\subsection{The response of the TA SD}

\begin{figure}[!htb]
    \centering
    \includegraphics[width=0.480\textwidth]{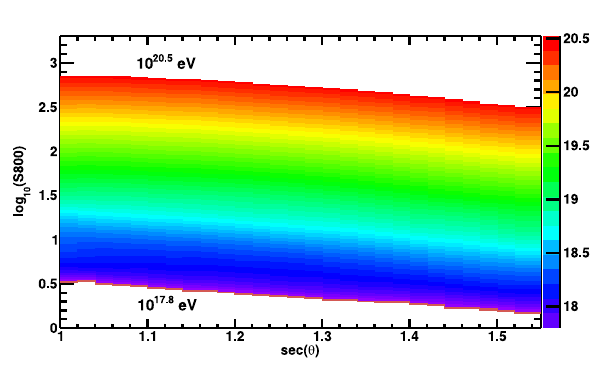}
    \caption{Energy calibration of EAS by readings of the TA SD~\cite{b:14}.}
    \label{f:6}
\end{figure}

\begin{figure}[!htb]
    \centering
    \includegraphics[width=0.480\textwidth]{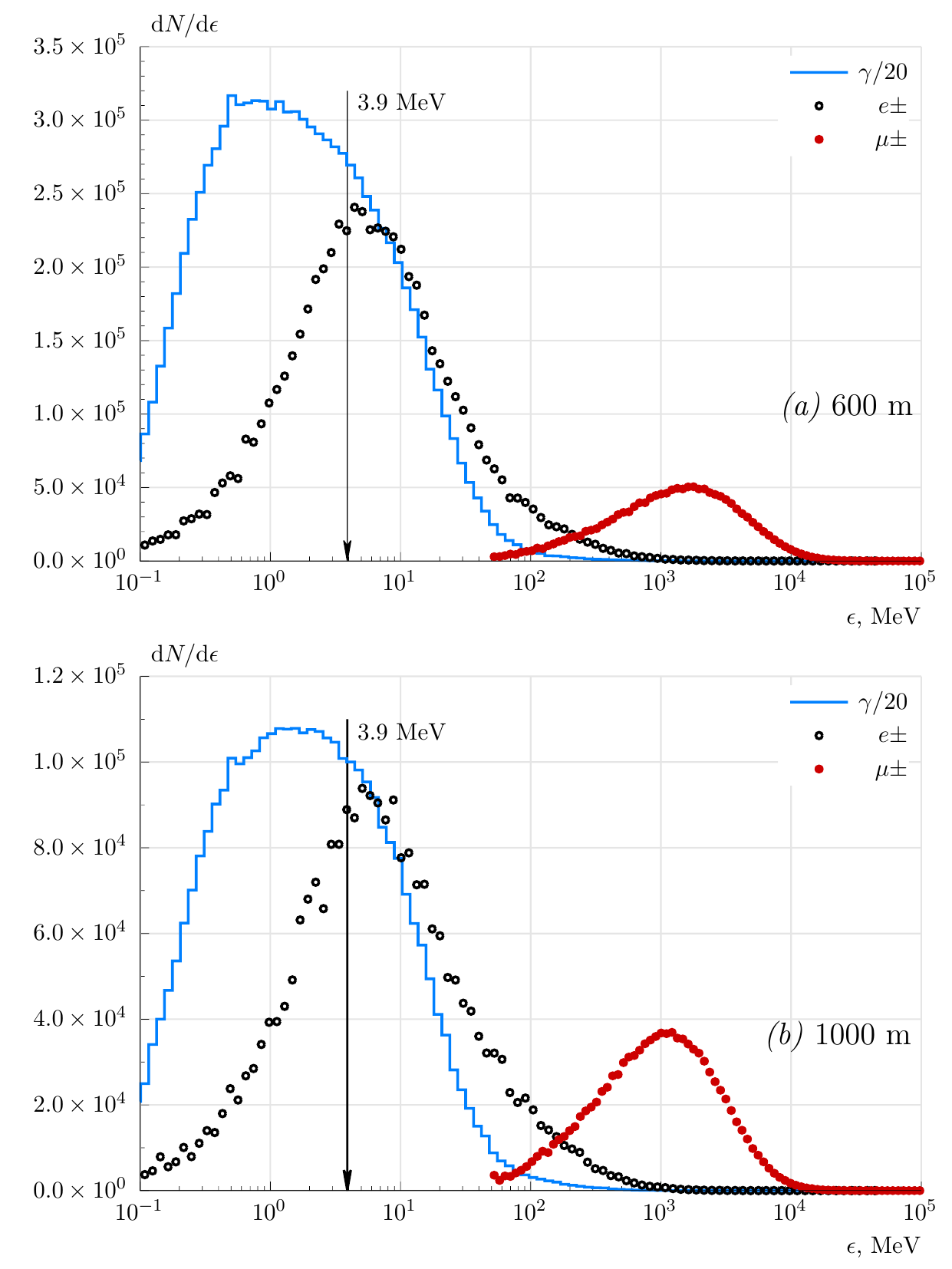}
    \caption{Energy spectra of EAS particles at different axis distances in showers with arrival direction $\cos\theta = 0.8$ with energy $10^{19}$~eV calculated within the framework of the \qgsii{} model for primary protons simulated for the conditions of TA experiment. The presented number of gamma-photons was scaled down by factor 20.}
    \label{f:7}
\end{figure}

In \fign{f:6} a nomogram is shown obtained in~\cite{b:14} for estimation of primary energy $\ESD$ from the readings of SD (\fign{f:1}). It is calculated with Monte Carlo method using the \corsika{} code~\cite{b:17} within the framework of the \qgsii{} hadron interaction model~\cite{b:18} for primary protons. Responses $S(800, \theta)$ were calculated using the Geant4 toolkit~\cite{b:15} for the entire detector area ($s = 3$~\sqrm). In inclined showers these values were measured in units:

\begin{equation}
    \VEM(\theta) = 2.05 \times \sec\theta~\text{[MeV],}
    \label{eq:6}
\end{equation}
which respect the growth of the charged particles energy deposit~\eqn{eq:2} in the medium due to increase of their track by factor $\sec\theta$. Unfortunately, some important details of the responses calculation in~\cite{b:14} were omitted. They could help understand the contradictions between the spectra presented in \fign{f:1}. To restore the necessary information we have performed independent calculations. The TA SD was simulated using Monte-Carlo method with a set of artificial air showers as a particle source. These showers were simulated using \corsika{} code~\cite{b:19} within the framework of the \qgsii{} model~\cite{b:17} for primary protons. Low-energy hadron interactions were treated with {\sc fluka} code~\cite{b:FLUKA}. In \fign{f:7} the spectra of shower particles at axis distances 600 and 1000~m are displayed. They reflect contributions of these particles to the total density recorded by the detector. Responses $S(800, \theta)$ were calculated using the custom code created at the Yakutsk array~\cite{b:19, b:20}. It has already been used previously (see, for example,~\cite{b:2, b:5}). During the calculation we assumed the following notions about the processes occurring inside the detector. All charged particles at axis distance $r$ with energy $\epsilon(r, \theta)$ which passed through the screen with a thickness

\begin{align}
    t\subscr(\theta) &= (2 \times 0.12 \times 7.874) \times \sec\theta = \nonumber \\
                     &\quad = 1.89 \times \sec\theta~\text{[\depth]}
    \label{eq:7}
\end{align}
with threshold energy

\begin{align}
    \epsilon\subscr(\theta) &= (1.89 \times 2.05) \times \sec\theta \approx \nonumber \\
                            &\quad \approx 3.9 \times \sec\theta~\text{[MeV],}
    \label{eq:8}
\end{align}
acquire a new energy:

\begin{equation}
    \varepsilon(r, \theta) = \epsilon(r, \theta) - \epsilon\subscr(\theta)\text{,}
    \label{eq:9}
\end{equation}
via partial ionization loss in the detector. To traverse the entire upper layer of the detector at angle $\theta$ a charged particle needs energy:

\begin{align}
    \vem(\theta) &= 2.05 \times 1.2 \times 1.032 \times \sec\theta \approx \nonumber \\
                 &\qquad \approx 2.54 \times \sec\theta~\text{[MeV].}
    \label{eq:10}
\end{align}
In such a case every particle with minimal energy

\begin{align}
    \epsilon\submin(\theta) &= \epsilon\subscr(\theta) + \vem(\theta) = \nonumber \\
                            &\quad = (3.9 + 2.54) \times \sec\theta~\text{[MeV],}
    \label{eq:11}
\end{align}
will yield a full-fledged single response. The number of such responses is proportional to the number of electrons and muons passing through the SD (see below). It can be seen from \fign{f:7} that not all electrons and a only a small portion of gamma-photons satisfy this condition. The rest with energy above the threshold \eqn{eq:8} will yield only ``cut'' responses, less than one. The shielding of a scintillator virtually does not affect the response of muons due to their high energies.

\subsection{Custom response generator code}

As mentioned above, a scintillation detector allows one to estimate the number of incident particles by measuring the energy deposited in the scintillator medium. Thus, for comparing experimental data and model calculations, it is necessary to use not straight values of the particle density $\rho(r)$ resulting from the simulation, but detector response. Also at axis distances $\sim 100 - 1000$~m the general particle flux is heavily dominated by gamma-photons; therefore their contribution to the resulting signal must be taken into account. Using the differential energy spectra of shower components at different axis distances (\fign{f:7}) one can create a model of a scintillation detector by taking into account various processes occurring during the passage of $m = 3$ types of particles that give the $u_m(\epsilon)$ response functions, which will determine the recorded densities of these particles.

Our model was created using the ``fast simulation'' approach~\cite{b:19, b:20}. Such models are well suited for processing data averaged over large datasets and are very fast. As a result, a one-dimensional model was created, which represents a scintillation detector as two layers of medium: a metal screen with threshold energy \eqn{eq:8} and a scintillator with energy deposit \eqn{eq:10} for a single response. Here, the three main recorded components were considered ($e\pm, \mu\pm$ and high-energy gamma-photons) and the most important physical processes associated with them occurring inside the detector:

\begin{itemize}
    \item $e\pm$: ionization, bremsstrahlung;
    \item $\mu\pm$: ionization;
    \item $\gamma$: pair production and recoil electrons from Compton scattering ($\delta e$).
\end{itemize}

Energy deposit inside the scintillator

\begin{equation}
    \Delta\epsilon_s(r, \theta) = n(r, \theta) \cdot \vem(\theta)
    \label{eq:12}
\end{equation}
is proportional to the number of particles $n(r, \theta)$ that passed through it and is measured in relative response units:

\begin{equation}
    \rho_s(r, \theta) = \Delta\epsilon_s(r, \theta) / \vem(\theta)\text{.}
    \label{eq:13}
\end{equation}
It is seen from \eqn{eq:12} and \eqn{eq:13} that for response unit \eqn{eq:10} only at the minimum of ionization curve (see \fign{f:8} below) the equality is satisfied:

\begin{equation}
    \rho_s(r, \theta) = n(r, \theta)\text{.}
    \label{eq:14}
\end{equation}
In all other cases $\rho_s(r, \theta) > n(r, \theta)$ and the total signal in detector will be a sum from all three components:

\begin{equation}
    \rho_s(r, \theta) = \rho_e(r, \theta)
    + \rho\submu(r, \theta)
    + \rho\subgam(r, \theta)\text{,}
    \label{eq:15}
\end{equation}
where contribution from each component of type $m$ is defined by the corresponding response function $\left<u_m(\epsilon, \theta)\right>$ and differential particle spectrum $I_m(\epsilon, r, \theta)$ (see \fign{f:7}) at given axis distance with respect to shower zenith angle $\theta$. Hence the spectrum provides a numerical description of a source function:

\begin{equation}
    \rho_m(r, \theta) = \sum_i \left<u_m(\epsilon_i, \theta)\right>
    \cdot
    I_m(\epsilon_i, r, \theta)\text{.}
    \label{eq:16}
\end{equation}
Note that the spectra presented in \fign{f:7} are given in the plane perpendicular to the shower axis. In other words, within the framework of calculations a shower nearly retains axial symmetry.

\subsection{The response of charged component}

\begin{figure}[!htb]
    \centering
    \includegraphics[width=0.480\textwidth]{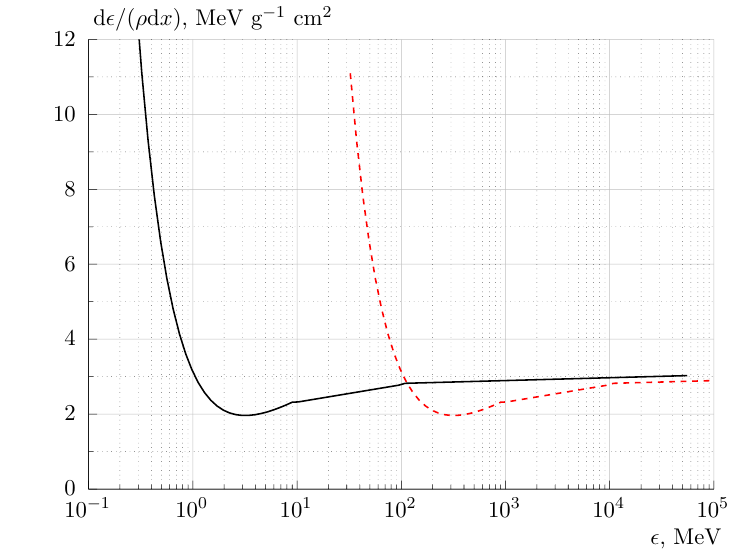}
    \caption{Differential ionization energy losses of electrons (solid curve) and muons (dashed curve) in water.}
    \label{f:8}
\end{figure}

\begin{figure}[!htb]
    \centering
    \includegraphics[width=0.480\textwidth]{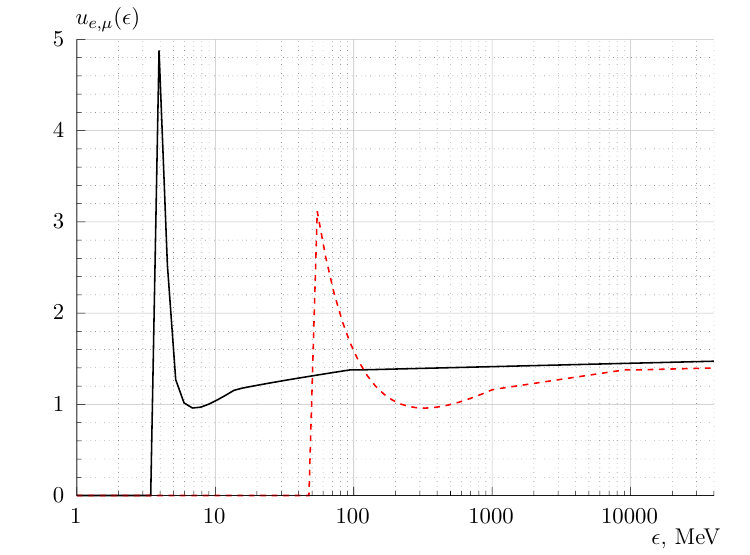}
    \caption{Responses of electrons (solid curve) and muons (dashed curve) in the TA scintillator.}
    \label{f:9}
\end{figure}

In \fign{f:8} the ionization losses of electrons and muons in water are shown. They are close to the same losses in a plastic scintillator. For electrons with energy $\epsilon \le 10$~MeV they can be represented as:

\begin{equation}
    \frac{{\rm d}\epsilon}{{\rm d}t} = 
    \frac{2.61}{\epsilon^{1.31}} +
    1.73 \cdot \lg \frac{2\epsilon}{10}~\edeposit\text{.}
    \label{eq:17}
\end{equation}
Within the region $10~\text{MeV} \le \epsilon \le 100~\text{MeV}$ they satisfy the condition:

\begin{equation}
    \frac{{\rm d}\epsilon}{{\rm d}t} = 
    2.38 + 0.464 \cdot \lg\frac{\epsilon}{10}~\edeposit\text{,}
    \label{eq:18}
\end{equation}
and at $\epsilon \ge 100$~MeV their rise much slower:

\begin{equation}
    \frac{{\rm d}\epsilon}{{\rm d}t} = 
    2.82 + 0.076 \cdot \lg\frac{\epsilon}{10}~\edeposit\text{.}
    \label{eq:19}
\end{equation}
When traversing the total thickness of a plastic $l = 1.2$~cm with a density $\rho = 1.032$~\dens{} in a shower with zenith angle $\theta$, a charge particle will deposit energy:

\begin{equation}
    \Delta \epsilon_1(\epsilon, \theta) = 
    \frac{{\rm d}\epsilon}{{\rm d}{t}} \times
    1.2 \times 1.03 \times \sec\theta~\text{[MeV].}
    \label{eq:20}
\end{equation}
In this case the electron response function can be expressed as:

\begin{equation}
    \left<u_e(\epsilon, \theta)\right> =
    \Delta \epsilon_1(\theta) / \vem(\theta)\text{,}
    \label{eq:21}
\end{equation}
shown in \fign{f:9} for a vertical shower. In inclined events the curve starts at threshold energy \eqn{eq:8} and shifts to the right. Equations \eqn{eq:17}-\eqn{eq:21} are also applicable to muons, in this case energy $\epsilon$ must be reduced by 100 times.

\subsection{The response of photon component}

\begin{figure}[!htb]
    \centering
    \includegraphics[width=0.480\textwidth]{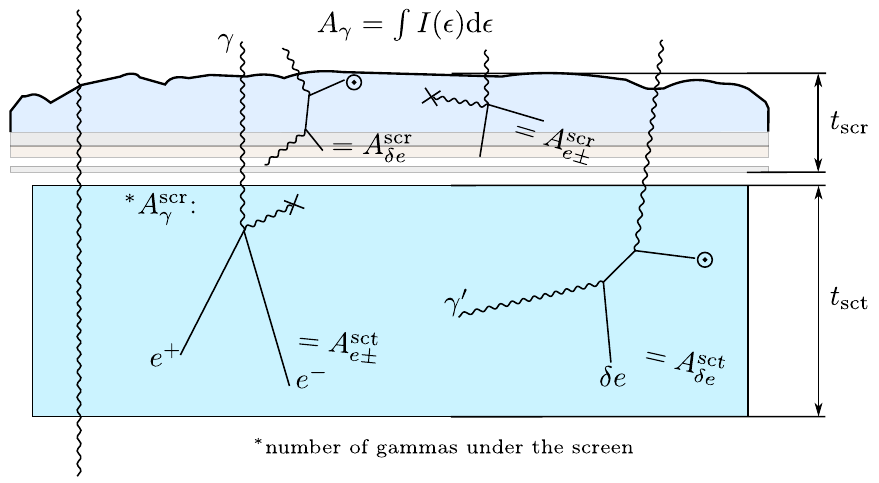}
    \caption{Passage of EAS photon component through the scintillation detector of the Yakutsk array~\cite{b:20}. The model considers the main processes resulting in generation of electrons/positrons that contribute to the total response.}
    \label{f:10}
\end{figure}

\begin{figure}[!htb]
    \centering
    \includegraphics[width=0.480\textwidth]{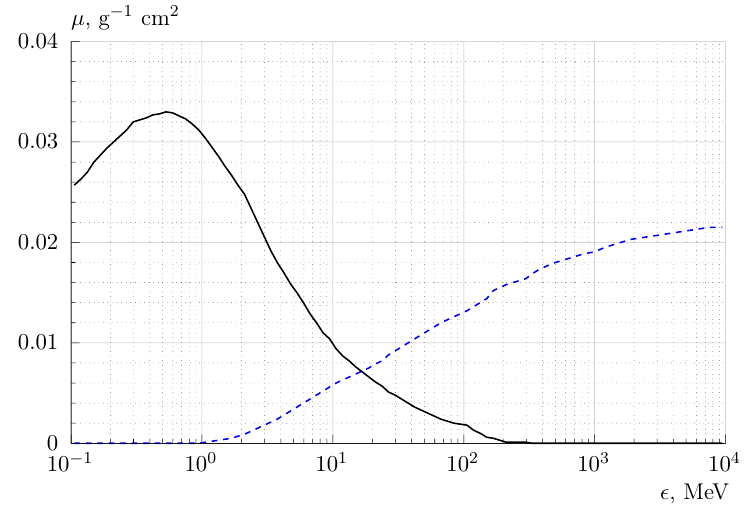}
    \caption{Mass coefficients of gamma attenuation in the medium due to generation of recoil electrons via Compton scattering (solid curve) and pair production (dashed curve).}
    \label{f:11}
\end{figure}

The overall picture of energy deposit from gamma-photons is presented in \fign{f:10} where SD of the Yakutsk array is shown as an example~\cite{b:20}. The multi-layered screen with the thickness $t\subscr = 2.5$~\depth{} consists of snow (on top), wood, plywood and aluminium. High-energy gamma-photons have two energy deposition channels that contribute to the response of a scintillation detector: pair production ($\gamma \rightarrow e^+ + e^-$) and recoil electrons arising from Compton scattering ($\delta e$):

\begin{equation}
    \rho\subgam = \rho_{e\pm} + \rho_{\delta e}\text{.}
    \label{eq:22}
\end{equation}
It's worth noting that interaction via each channel may occur inside both detector roof (scr) and scintillator medium (sct):

\begin{equation}
    \left.
    \begin{aligned}
        \rho\subpair  =  \rho\subpair\ovrscr + \rho\subpair\ovrsct\text{,} \\
        \rho\subcomp  =  \rho\subcomp\ovrscr + \rho\subcomp\ovrsct\text{.}
    \end{aligned}
    \right.
    \label{eq:23}
\end{equation}
Here the measured response from gamma-photons depends on the number $I(\varepsilon)$ of interactions in the primary stream $I_0(\varepsilon)$ (see \fign{f:7}) passing through the layers of mediums that form the SD with thickness $t$:

\begin{equation}
    \Delta I_{\gamma}(\varepsilon) =
    I_0(\varepsilon) \times (1 - e^{-\mu(\varepsilon) \cdot t})\text{.}
    \label{eq:24}
\end{equation}
Mass coefficients of interaction with the medium are shown in \fign{f:11}.

\begin{figure}[!htb]
    \centering
    \includegraphics[width=0.480\textwidth]{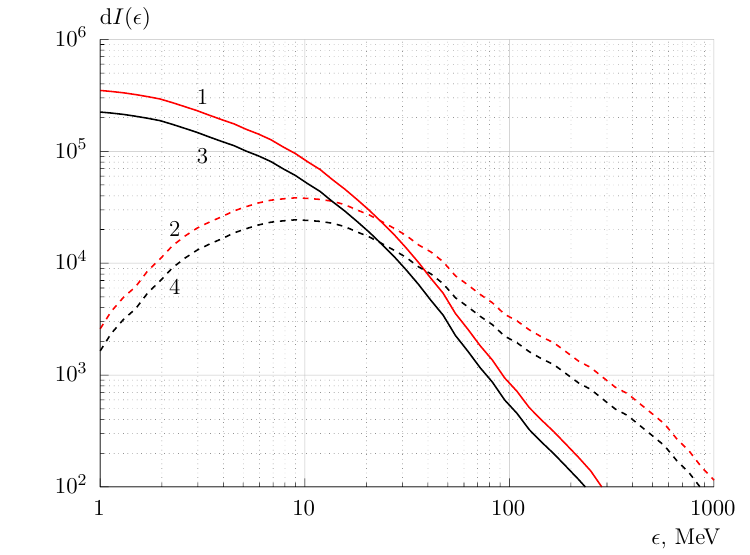}
    \caption{The number of gamma-photons at axis distance $r = 600$~m in vertical showers with $\E = 10^{19}$~eV from primary protons calculated using the \qgsii{} model for the TA: solid lines~--- Compton scattering; dashed lines~--- pair production; 1 and 2~--- processes occurring inside the screen, 3 and 4~--- inside the scintillator.}
    \label{f:12}
\end{figure}

As an example in \fign{f:12} the numbers of gamma-photons \eqn{eq:24} are shown at axis distance $r = 600$~m in vertical showers initiated by primary protons with energy of $10^{19}$~eV calculated within the framework of the \qgsii{} model for conditions of TA. It is seen that inside the screen occurs $1.89/(1.2 \times 1.032) \approx 1.53$ times more interactions than inside the scintillator.

\begin{figure}[!htb]
    \centering
    \includegraphics[width=0.480\textwidth]{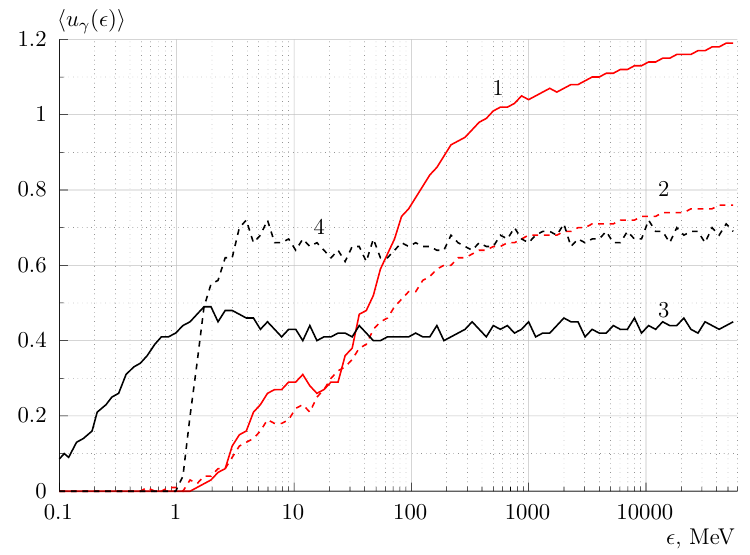}
    \caption{Response functions of photon component in vertical EAS: solid curves~--- Compton scattering of electrons; dashed curves~--- pair production; 1 and 2~--- processes occurring inside the screen, 3 and 4~--- inside the scintillator.}
    \label{f:13}
\end{figure}

The response function of gamma-photons consists of two components. First of them, $u\subcomp(\epsilon)$, refers to recoil electrons  from Compton scattering. As a result of this process, a separate photon (\fign{f:7}) can produce an electron with energy ranging from 0 to input value $\epsilon$. Its production is equally probable at any depth of screen or scintillator. At the same time, the ionization losses of electron still satisfy the equations \eqn{eq:17}-\eqn{eq:21}, and random values of its energy and depth of origin are taken into account in the response 

\begin{equation}
    \left<u\subcomp(\epsilon,\theta)\right> =
    \left<\Delta \epsilon_1(\theta)\right> / \vem(\theta)\text{,}
    \label{eq:25}
\end{equation}
where average losses $\left<\Delta \epsilon_1(\theta)\right>$ were determined via Monte-Carlo method by 500-fold drawing (\fign{f:13}). The response of second component $\left<u\subcomp(\epsilon, \theta)\right>$ was obtained with the same technique.

\subsection{Lateral distribution of EAS particles}

\begin{figure}[!htb]
    \centering
    \includegraphics[width=0.480\textwidth]{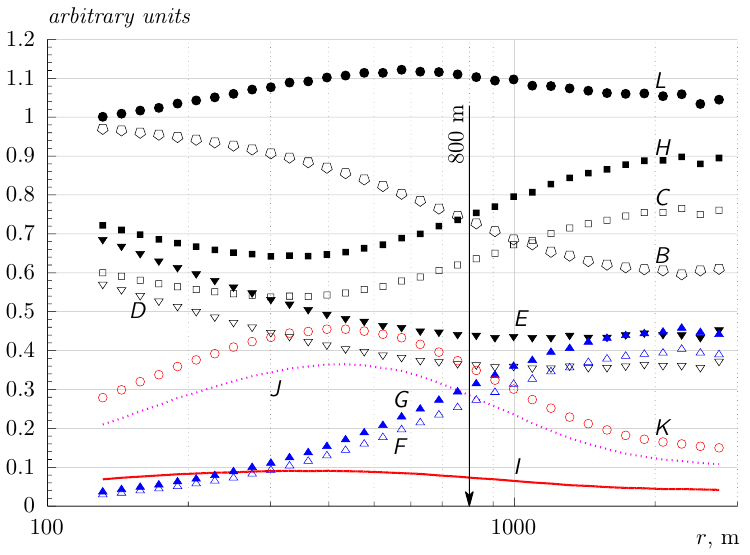}
    \caption{Lateral distributions of particles in vertical showers initiated by primary protons with energy $10^{19}$~eV. All values are normalized to the density \eqn{eq:26}. Calculations were performed using the \qgsii{} model for the conditions of TA.}
    \label{f:14}
\end{figure}

In \fign{f:14} the lateral distributions are shown of three components in vertical EAS with $\E = 10^{19}$~eV initiated by primary protons in the upper layer of the TA SD (\fign{f:2}). For clarity: all data were normalized to the total LDF of charged particles constituted from electrons with threshold energy $\ElThr = 1$~MeV and muons with threshold energy $\MuThr = 50$~MeV:

\begin{equation}
    \rho_{e+\mu}(\ElThr = 1~\text{MeV}, \MuThr = 50~\text{MeV}) =
    \rho\subch(r)\text{.}
    \label{eq:26}
\end{equation}
The 50~MeV value is the minimum possible energy of muons in \corsika{}. Relative contributions from electrons

\begin{equation}
    \rho_e\ovrRelUn(r) = 
    \frac{\rho_e(\varepsilon\subthr = 1~\text{MeV}, r)}{\rho\subch(r)}
    \label{eq:27}
\end{equation}
and muons

\begin{equation}
    \rho\submu\ovrRelUn(r) = 
    \frac{\rho\submu(\varepsilon\subthr = 50~\text{MeV}, r)}{\rho\subch(r)}
    \label{eq:28}
\end{equation}
are shown with curves $B$ and $F$ correspondingly. The sum of \eqn{eq:27} and \eqn{eq:28} equals to 1. The $D$ curve represents electrons with threshold energy $\ElThr = 3.9$~MeV:

\begin{equation}
    \rho_e\ovrRelUn(r) =
    \frac{\rho_e(\varepsilon\subthr = 3.9~\text{MeV}, r)}{\rho\subch(r)}\text{,}
    \label{eq:29}
\end{equation}
which can pass the entire thickness of the screen and reach the scintillator. The $C$ curve represents the sum of \eqn{eq:28} and \eqn{eq:29}. Essentially, it is the density that would be recorded by a Geiger-M\"{u}ller counter.

A scintillation detector, due to the processes described above, reacts to these particles differently. The $E$ curve represents the responses from electrons with $\ElThr = 3.9$~MeV threshold:

\begin{equation}
    s_e\ovrRelUn(r) =
    \frac{s_e(\varepsilon\subthr = 3.9~\text{MeV}, r)}{\rho\subch(r)}\text{.}
    \label{eq:30}
\end{equation}
The $G$ curve reflects the responses from muons with $\MuThr = 50$~MeV:

\begin{equation}
    s\submu\ovrRelUn(r) =
    \frac{s\submu(\varepsilon\subthr = 50~\text{MeV})}{\rho\subch(r)}\text{.}
    \label{eq:31}
\end{equation}
The $H$ curve represents the sum of \eqn{eq:30} and \eqn{eq:31}. It is the response from cascade electrons and muons that would be recorded by TA SD. It seen that its LDF lies $\approx 1.175$ times higher than the LDF of direct number of these particles (the $C$ curve) in the entire range of axis distances $r$.

The $J$ and $I$ curves represent the responses, correspondingly, from photons in the scintillator and detector screen. It is seen that the former is significantly higher than the later: at $r = 800$~m the difference amounts to times 4. The $K$ curve refers to the sum of $J$ and $I$:

\begin{equation}
    s\ovrRelUn\subgam(r) = \frac{s\subgam(r)}{\rho\subch(r)}\text{.}
    \label{eq:32}
\end{equation}

And finally, the $L$ curve represents the sum of responses \eqn{eq:30}, \eqn{eq:31} and \eqn{eq:32} which at $r = 800$~m is greater than the sum of \eqn{eq:30} and \eqn{eq:31} by factor $1.11 / 0.74 \approx 1.5$. This shows that the contribution of photonic component in the total TA SD signal is significant.

\subsection{Zenith-angular dependencies of response}

\begin{figure}[!htb]
    \centering
    \includegraphics[width=0.480\textwidth]{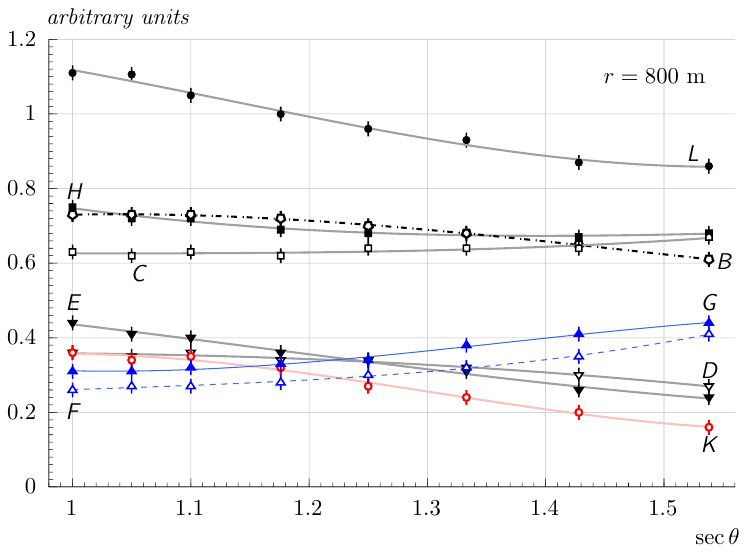}
    \caption{Zenith-angular dependencies of responses from components which constitute the $S(800,\theta)$ signal calculated using the \qgsii{} model for primary protons. Symbols are the same as in \fign{f:14}, lines were drawn for easier points tracking.}
    \label{f:15}
\end{figure}

In \fign{f:15} the zenith-angular dependencies of signals from EAS components are shown at axis distance $r = 800$~m. All designations are the same as in \fign{f:14}. It is seen that the sum of responses \eqn{eq:30} and \eqn{eq:31} (the $H$ curve) with increase of showers zenith angles gradually decreases and approaches to the sum of electrons and muons (the $C$ curve), which virtually does not change. This suggests that the choice of a response unit in the form of \eqn{eq:10} is correct. It is worth noting that all three kinds of EAS particles are closest to each other at $\sec \theta \approx 1.25$ (see below $ \theta \approx 38\degr$) and contribution from photons (the $K$ curve) in the total response from all particles (the $L$ curve) is $\approx 28$\%, i.e. close to the discrepancy between energy calibrations of TA (\fign{f:1}).

\begin{figure}[!htb]
    \centering
    \includegraphics[width=0.480\textwidth]{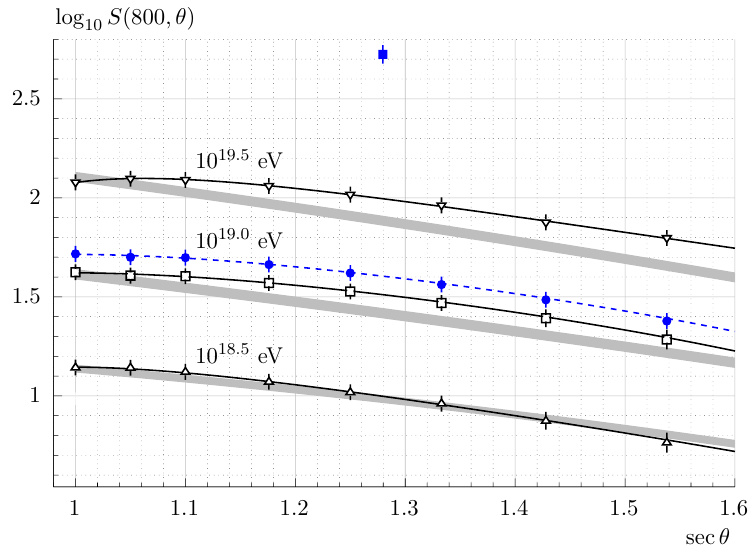}
    \caption{Zenith-angular dependencies of $S(800, \theta)$ calculated according to the \qgsii{} model for primary protons: gray bands refer to the nomogram in \fign{f:6}~\cite{b:14} with selected primary energies. Dark circles~--- calculation with response unit~\eqn{eq:6}; light symbols~--- responses calculated in units~\eqn{eq:10}. Dark square represents the giant TA event~\cite{b:21}.}
    \label{f:16}
\end{figure}

In \fign{f:16} zenith-angular dependences are shown of the total response $S(800, \theta)$ collected from the entire area of TA SD ($s = 3$~\sqrm). Gray bands refer to EAS energies considered in \fign{f:6}. Their width reflects the accuracy of the transfer of these data into \fign{f:16}. Errors are determined by methodological inaccuracies of our calculations. Dark circles represent calculation with response unit \eqn{eq:6}, light symbols~--- with response unit \eqn{eq:10}. Square denotes the Amaterasu event registered by TA~\cite{b:21} (see Section~4.3). It can be seen that at energy $~\approx 3.16 \times 10^{18}$~eV our calculations and the results given in~\cite{b:14} are consistent with each other in the entire considered range of zenith angles. In other cases, there is no such agreement. Moreover, calculations with response unit \eqn{eq:6} (dashed line) used in~\cite{b:14} differ in absolute value from calculations with response unit \eqn{eq:10} by factor $\approx 1.27$. This difference remains at other energies.

\section{Results and discussion}

\subsection{Estimation of $\ESD$}

It is seen from \fign{f:16} that for vertical showers the nomorgam~\cite{b:14} agrees with our calculations. They can be expressed with a formula:

\begin{equation}
    \ESD = \ESDOne \times \rho_s(800, 0\degr)^{1.025 \pm 0.010}~\text{[eV],}
    \label{eq:33}
\end{equation}

The corresponding proportional coefficient $\ESDOne$ and classification parameter $\rho_s(r, \theta)$ are given in the Table~\ref{t:1} below. The $\ESDOne$ value equals to the energy of a vertical shower with the response density $\rho_s(800, 0\degr) = 1$~\pdens{} measured at axis distance $r = 800$~m. Essentially, experimentally primary energy is estimated in units of the energy of some reference EAS event. Column~2 lists elevation of array above sea level, column~4~--- spectra cutoff threshold of EAS particles, below which the events were not taken into account during calculations (\fign{f:7}). Column~5 shows the zenith angle, to which the density $\rho_s(r, \theta)$ was recalculated. Row~2 lists primary energy estimation according to the formula \eqn{eq:33} with response unit \eqn{eq:6}.

\begin{table*}[!htb]
    \centering
    \caption{Proportional coefficients $\ESDOne$ and classification parameters $\rho_s(r, \theta)$ in the formula \eqn{eq:33} which were either measured in experiment (lines 1, 3, 5, 7) or obtained in \qgsii, $p$ (column 9); rows and columns were numbered for convenience}
    \label{t:1}
    \setlength{\tabcolsep}{0.5em}     
    \renewcommand{\arraystretch}{1.2} 
    \newcommand{\CTwo}{0.04\textwidth}
    \newcommand{\CThree}{0.04\textwidth}
    \newcommand{\CFour}{0.06\textwidth}
    \newcommand{\CFive}{0.02\textwidth}
    \newcommand{\CSeven}{0.11\textwidth}
    \newcommand{\CEight}{0.11\textwidth}
    \newcommand{\CNine}{0.14\textwidth}
    \begin{tabular}{%
        r
        R{\CTwo}
        R{\CThree}
        C{\CFour}
        R{\CFive}
        R{\CSeven}
        R{\CEight}
        L{\CNine}
        l
        }
        \hline
        \hline
        1                     & \multicolumn{1}{c}{2} & \multicolumn{1}{c}{3} &
        \multicolumn{1}{c}{4} & \multicolumn{1}{c}{5} & \multicolumn{1}{c}{6} &
        \multicolumn{1}{c}{7} & \multicolumn{1}{c}{8} & \multicolumn{1}{c}{9} \\
        \hline
        %
        & \multicolumn{1}{C{\CTwo}}{$\xobs$, \depth}
        & \multicolumn{1}{C{\CThree}}{$r$, m}
        & cut-off, MeV
        & \multicolumn{1}{C{\CFive}}{$\theta$}
        & \multicolumn{1}{C{\CSeven}}{$\rho_s(r,\theta)$, \pdens}
        & \multicolumn{1}{C{\CEight}}{$\ESDOne \times 10^{17}$}
        & \multicolumn{1}{C{\CNine}}{Source, $10^{19}$~eV}
        & Array [reference] \\
        \hline
        1 &  876 &  800 & $\ge 3.9$ &  $0\degr$ & $13.28 \pm 0.93$ & $7.06 \pm 0.25$ & 
            \qgsii, $p$ & TA [this work] \\
        2 &  876 &  800 & $\ge 3.9$ &  $0\degr$ & $16.87 \pm 1.18$ & $5.52 \pm 0.20$ &
            \qgsii, $p$ & TA [this work] \\
        3 & 1020 &  600 & $\ge 6.0$ &  $0\degr$ & $24.60 \pm 0.30$ & $3.76 \pm 0.30$ &
            experiment & Yakutsk \cite{b:2} \\
        4 & 1020 &  600 & $\ge 6.0$ &  $0\degr$ & $25.40 \pm 0.02$ & $3.64 \pm 0.02$ &
            \qgsii, $p$ & Yakutsk \cite{b:2} \\
        5 & 1016 &  600 & $\ge 9.0$ &  $0\degr$ & $13.60 \pm 0.20$ & $7.04 \pm 0.56$ & 
            experiment & HP \cite{b:22, b:23} \\
        6 & 1016 &  600 & $\ge 9.0$ & $38\degr$ & $13.80 \pm 0.02$ & $6.90 \pm 0.02$ &
            \qgsii, $p$ & HP \cite{b:4} \\
        7 &  875 & 1000 & $\ge 9.0$ & $38\degr$ & $5.00 \pm 0.02$ & $19.00 \pm 0.05$ &
            experiment & Auger \cite{b:23} \\
        8 &  875 & 1000 & $\ge 9.0$ & $38\degr$ & $5.02 \pm 0.02$ & $19.13 \pm 0.02$ &
            \qgsii, $p$ & Auger \cite{b:4} \\
        9 &  876 & 1000 & $\ge 3.9$ & $38\degr$ & $4.90 \pm 0.16$ & $19.60 \pm 0.60$ &
            \qgsii, $p$ & TA [this work] \\
        10 & 876 & 1000 & $\ge 3.9$ & $38\degr$ & $6.10 \pm 0.15$ & $15.60 \pm 0.50$ &
            \qgsii, $p$ & TA [this work] \\
        11 & 876 & 1000 & $\ge 3.9$ &  $0\degr$ & $5.80 \pm 0.15$ & $16.50 \pm 0.50$ &
            \qgsii, $p$ & TA [this work] \\
        \hline
        \hline
    \end{tabular}
\end{table*}

Before continuing the analysis of the TA energy estimation, let's consider some results~\cite{b:4} that would be useful in the further discussion. Rows~3 and 4 of the Table~\ref{t:1} list the parameters of formula \eqn{eq:33} which were either measured experimentally or calculated for the Yakutsk array~\cite{b:2}. They are consistent with each other within 3\% thus suggesting that simulation is adequate to experiment and vice versa. Using this simple model we have performed ``diagnostics'' of other arrays of our interest at that moment. Row~5 presents the experimental data of Haverah Park (HP)~\cite{b:22, b:23}. This array is located at atmospheric depth $x = 1016$~\depth, close to the level of the Yakutsk array. Both use the same classification parameter, particle density $\rho_s(600, 0\degr)$ determined with the formula:

\begin{equation}
    \rho_s(600, 0\degr) = \rho_s(600, \theta) \cdot
    \exp\frac{(\sec\theta - 1) \cdot x}{\lambda}~\text{[\pdens]}
    \label{eq:34}
\end{equation}
with attenuation lengths $\lambda = 500 \pm 30$~\depth~\cite{b:2} and $760 \pm 40$~\depth~\cite{b:23} correspondingly. Row~6 lists the results of calculations performed for HP where measured densities $\rho_s(600, \theta)$ were recalculated not to vertical, but to the value $\left<\theta\right> = 38\degr$. Only in this case observations~\cite{b:22, b:23} are consistent with calculations~\cite{b:4}.

\begin{figure}[!htb]
    \centering
    \includegraphics[width=0.480\textwidth]{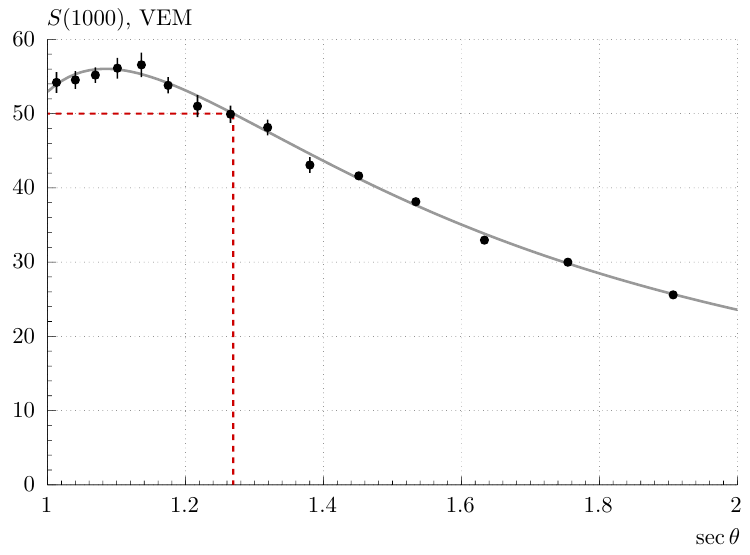}
    \caption{The attenuation curve \eqn{eq:36} used for recalculation of the measured response $S(1000, \theta)$ to the angle $\left<\theta\right> = 38\degr$. The parameter $S_{38} \approx 50$~VEM (dashed line) corresponds to primary energy $1.05 \times 10^{19}$~eV~\cite{b:24}.}
    \label{f:17}
\end{figure}

\begin{figure}[!htb]
    \centering
    \includegraphics[width=0.480\textwidth]{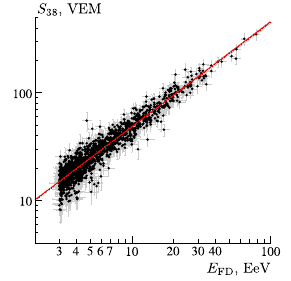}
    \caption{Correlation between $S_{38}$ and $\EFD$ (see Fig.~41 in~\cite{b:24}).}
    \label{f:18}
\end{figure}

Row~7 lists parameters of the relation \eqn{eq:33} for Auger obtained in \cite{b:4} from the fundamental formula~\cite{b:24}:

\begin{equation}
    \ESD = A \times \left[
        \frac{S(1000, \theta)}{\fcic(\theta) / \VEM}
    \right]^{1.025 \pm 0.007}~\text{[eV],}
    \label{eq:35}
\end{equation}
where $A = (1.90 \pm 0.05) \times 10^{17}$~eV. The $S(1000, \theta)$ parameter is equal to the total response of all shower particles in SD with 3.6~m diameter ($s_{\text{SD}} = 10.2$~\sqrm) in units $\VEM(0\degr) \approx 270$~MeV. The $\ESD$ value was determined from responses $S(1000, \theta)$ recalculated to $\left<\theta\right> = 38\degr$ using the attenuation curve (Fig.~\ref{f:17}):

\begin{equation}
    \fcic(\theta) = 1 + ax + bx^2 + cx^3\text{,}
    \label{eq:36}
\end{equation}
where $x = \cos^2(\theta) - \cos^2(\left<\theta\right>)$, $a = 0.980 \pm 0.004$, $b = -1.68 \pm 0.01$ and $c = -1.30 \pm 0.45$. At the same time the Auger experiment used another parameter $S_{38} \equiv S(1000, \theta)/\fcic(\theta)$ in the following relation:

\begin{equation}
    \EFD = A \times S_{38}^{1.025 \pm 0.007}~\text{[eV],}
    \label{eq:37}
\end{equation}
used for energy estimation by readings of FDs registering the EAS fluorescent light emission. The correlation between these two parameters is shown on \fign{f:18}.

\begin{figure}[!htb]
    \centering
    \includegraphics[width=0.480\textwidth]{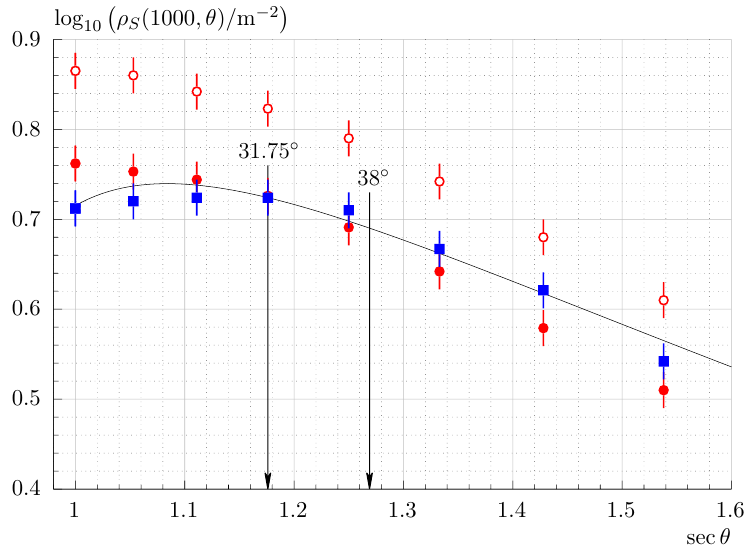}
    \caption{Zenith-angular dependencies of $\rho_s(1000, \theta)$ obtained within the framework of the \qgsii{} model for primary protons with energy $10^{19}$~eV. Solid curve refers to \fign{f:17}. Dark circles~--- calculation of response with units~\eqn{eq:10}, light circles~--- calculation with units~\eqn{eq:6}. Squares represent the result of calculations~\cite{b:4} for Yakutsk scintillation detector as if it was placed side by side with TA's or Auger's SD.}
    \label{f:19}
\end{figure}

Squares in \fign{f:19} represent the results of calculation of the response of Yakutsk SD~\cite{b:4} (\fign{f:3}) as if it was placed next to Auger's or TA's SDs at corresponding elevations above sea level. The $\rho_s(1000, \theta)$ density is normalized to the detector's unit of area. Parameters of formula \eqn{eq:37} that correspond to calculations~\cite{b:4} are listed in row~8. They are identical to the results~\cite{b:24} in row~7. At the same time they do not contradict the calculations for TA with response unit~\eqn{eq:10} (dark circles). The attenuation curves shown on \fign{f:19} intersect at $\theta \approx 32\degr$. Energy estimations derived from the $S_{38}$ parameter using the formula~\eqn{eq:37} virtually coincide (rows 7, 8 and 9). This suggests that if a chosen response unit was physically sound, then differently designed SDs should give similar response LDFs.

Empty circles in \fign{f:19} represent the results of $\rho_s(1000, \theta)$ calculation in units \eqn{eq:6}. They are lying $\approx 1.26$ times higher than other data. At $\theta = 38\degr$ they give the parameters of formula~\eqn{eq:37} (row~10), which relate with their analogues in row~9 as:

\begin{align}
    \ESDTen &= (19.6 / 15.6) \times \ESDTen \approx \nonumber\\
            &\quad \approx 1.26 \times \ESDSix\text{,}
    \label{eq:38}
\end{align}
where indices 6 and 10 denote response units \eqn{eq:6} and \eqn{eq:10} correspondingly. Now let's consider a quote from work~\cite{b:14} (page~101)~\footnote{The numbering of formulas corresponds to this paper.}:

\begin{quotation}
    ``\textbf{5.6 Energy Scale.} To reduce the model dependence of the TA SD energy scale, the energy values obtained from the energy estimation table (Figure~5.5) are calibrated against the TA fluorescence detector using events that are seen in common by both TA SD and FD and are well reconstructed by each detector separately. In order to match the TA FD energy, the TA SD energies determined from the energy estimation table (Figure 5.5) need to be reduced by a factor 0.787. In other words, when the 102 energy estimation procedure derived from the \corsika{} surface detector Monte-Carlo is applied to the real data, the predicted event energies are on average 27\% higher than those of the fluorescence detector:

\begin{equation}
    \ESDCor = 1.27 \times \EFD
    \label{eq:39}
\end{equation}
Figure~5.6a shows the energy of the TA SD plotted versus the energy of the TA FD, after the TA SD energy has been reduced by a factor of 1.27.''
\end{quotation}
From this piece, in the context with the above, the following interpretation is possible. Relations~\eqn{eq:38} and~\eqn{eq:39} are structurally the same. Their left sides are related to  primary energy estimations $\ESDCor$ and $\ESDTen$ which follow from readings of SD. Values on the right are $1.26-1.27$ times lower than the former. One of them ($\EFD$) is the estimation of EAS energy obtained using FD~\cite{b:14}.

The second value is the result of our calculation of $\ESDSix$ energy in the formula~\eqn{eq:1} with $\ESDOne = (1.81 \pm 0.08) \times 10^{17}$~eV. It leads to an erronous result due to the incorrectly selected response unit~\eqn{eq:6} in~\cite{b:14}. From a physical point of view it is obvious that a muon (or electron) with such energy would not be able to traverse the entire thickness of a scintillator (see \fign{f:2}). For this, the particle energy must be

\begin{equation}
    \frac{\vem(\theta)}{\VEM(\theta)} = 
    \frac{
        2.05 \times 1.2 \times 1.032 \times \sec\theta
    }{
        2.05 \times \sec\theta
    } \approx 1.24
    \label{eq:40}
\end{equation}
times higher. It is not entirely clear how this fact is connected with formal equality $\ESDSix \approx \EFD$ in expressions~\eqn{eq:38} and \eqn{eq:39}.

\subsection{Estimation of $\EFD$}

\begin{figure}[!htb]
    \centering
    \includegraphics[width=0.480\textwidth]{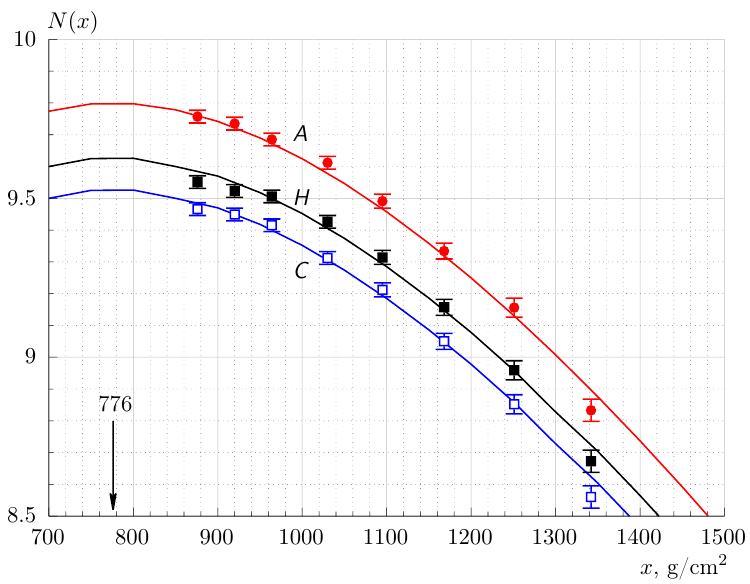}
    \caption{Cascade curves of EAS charged particles calculated within the framework of the \qgsii{} model for primary protons with energy $10^{19}$~eV. $A$~--- function~\eqn{eq:41}. Amplitudes of curves $H$ and $C$ are less then the amplitude of $A$ by 1.48 and 1.87 times correspondingly. Their designations are identical to \fign{f:14}. Symbols represent integrals~\eqn{eq:42} of densities~\eqn{eq:26}, \eqn{eq:30} + \eqn{eq:31} and~\eqn{eq:28} + \eqn{eq:29}, respectively.}
    \label{f:20}
\end{figure}

In order to understand this, we have attempted to estimate the $\EFD$ independently. Curve $A$ in \fign{f:20} represents the longitudinal profile $N(x)$ of the full number of charged particles (electrons and muons) as a function of the distance $x - x_0$ between the thickness of atmosphere $x = 876 \times \sec\theta$~\depth{} above the array and the point of first interaction $x_0 = 37$~\depth{} of a proton with energy $10^{19}$~eV, obtained within the framework of the \qgsii{} model. The maximum of the curve is equal to $\xmax = 776 \pm 7$~\depth{}, the number of particles in the maximum $\Nmax = 6.3 \times 10^9$. Analytically the cascade curve is expressed with the Gaisser-Hillas function~\cite{b:25}:

\begin{align}
    N(x) &= \Nmax \times \nonumber \\
         &\quad\times
          \left(
              \frac{x - x_0}{\xmax - x_0}
          \right)^{\frac{\xmax - x_0}{\lambda}} \times \nonumber \\
         &\quad\quad \times \exp\frac{\xmax - x}{\lambda}~\text{,}
    \label{eq:41}
\end{align}
where $\lambda$ is a phenomenological parameter $\sim 70$~\depth. It was used in~\cite{b:14} for estimation of $\EFD$. Symbols on \fign{f:20} represent summary values

\begin{equation}
    N(x) = 2\pi \int_0^{r*} \rho(r, \theta){\rm d}r\text{,}
    \label{eq:42}
\end{equation}
of the densities shown in~\fign{f:14}. They were calculated in the range of axis distances from 0 to $r* = 10^4$~m using the LDF~\eqn{eq:43} measured at the Yakutsk array~\cite{b:19, b:20}, and turned out to be quite acceptable the analytical representation of the TA data:

\begin{align}
    \rho(r, \theta) &= \rho_s(600, \theta) \cdot
    \left(
        \frac{r + r_1}{600 + r_1}
    \right)^{2} \times \nonumber \\
    &\quad \times
    \left(
        \frac{r + r_0}{600 + r_0}
    \right)^{\beta - 2} \cdot
    \left(
        \frac{r + r_2}{600 + r_2}
    \right)^{11}\text{,}
    \label{eq:43}
\end{align}
where $r_0 = 8$, $r_1 = 10$, $r_2 = 10^4$ and $\beta(\theta)$ and $\rho_s(600, \theta)$ are free parameters obtained during the $\chi^2$-minimization. Light squares represent summary densities~\eqn{eq:28} and \eqn{eq:29}. They reflect full number of muons and electrons with threshold energy $\varepsilon\subthr = 3.9$~MeV at observation level $x$. Dark squares reflect the sum of their responses \eqn{eq:30} and \eqn{eq:31}. It is seen that the values~\eqn{eq:42} agree with cascade curve~\eqn{eq:41} measured with FD. It means that the \qgsii{} model reflects experimental data quite adequately and can serve as a reliable basis for description of many processes occurring during the EAS development.

The full integral of the cascade curve \eqn{eq:41} can be represented as the sum:

\begin{align}
    I &= \int_0^{\xobs} N(x){\rm d}x +
         \int^{\infty}_{\xobs} N(x){\rm d}x \approx\nonumber \\
      &\quad\approx 575.4~\text{\depth} \times \Nmax \approx
                    3.63 \times 10^{12}~\text{\depth.}
    \label{eq:44}
\end{align}
The first term here is the summary path of all charged particles in the atmosphere down to observation level $\xobs = 876 \times \sec\theta$~\depth; second one~--- summary path of survived particles in the ground until their complete stopping. Absolute value of $I$ does not depend on the zenith angle of EAS arrival, since change of zenith angle only results in redistribution of values of constituents.

The cascade curve~\eqn{eq:41} was measured experimentally using FD. It allows one to estimate the $\EFD$ energy in units of VEM described in~\cite{b:14}:

\begin{align}
    \EFD &= \VEM \times I = \nonumber \\
         &\quad= 2.05 \times 10^6 \cdot 3.63 \times 10^{12} = \nonumber \\
         &\quad\quad = 7.44 \times 10^{18}~\text{eV.}
    \label{eq:45}
\end{align}
To account the energy fraction not associated with electromagnetic component of EAS, its value must be increased by $\approx 7$\%. Then the final value is:

\begin{equation}
    E_{\text{FD}, 6} = 1.07 \times 7.44 \times 10^{18} \approx 7.96 \times 10^{18}~\text{eV.}
    \label{eq:46}
\end{equation}
Here index 6 hints that the equality mentioned in Section~4.1 is finally satisfied:

\begin{equation}
    E_{\text{FD}, 6} \approx \ESDSix\text{.}
    \label{eq:47}
\end{equation}
But this is the minimum possible estimation of primary energy. As it follows from~\fign{f:8} and \fign{f:9}, on both sides of the minimum of ionization curve energy losses of electrons and muons exceed 1~VEM. These losses are especially high just before stopping of particle, when they increase tenfold. Therefore, estimation \eqn{eq:46} should be additionally increased by $\delta$ times. The ionization losses of electrons and muons in the atmosphere and in ground are no different from the energy deposit in the SD scintillator. In our case, this is confirmed by cascade curves $C$ and $H$ in \fign{f:20}, which are smaller than the $A$ curve in absolute value by 1.87 and 1.48 times respectively. The value of their difference between each other~--- $\delta = 1.87/1.48 = 1.264$~--- is just the desired adjustment.

The final value of $\EFD$ is:

\begin{align}
    \EFD &= \delta \times E_{\text{SD}, 6} = \nonumber \\
         &\quad = 1.264 \times 7.960 \times 10^{18}~\text{eV} \approx \nonumber \\
         &\quad\quad \approx 10^{19}~\text{eV.}
    \label{eq:48}
\end{align}

Let us pay attention to an amazing coincidence: if one applies the response unit~\eqn{eq:10} to expression~\eqn{eq:45}, then the resulting estimation is:

\begin{align}
    E_{\text{FD}, 10} &= 1.07 \times \vem(0\degr) \times I = \nonumber \\
                      &\quad = 1.07 \times 2.54 \times 3.63 \times 10^{12} = \nonumber \\
                      &\qquad = 9.87 \times 10^{18}~\text{eV,}
    \label{eq:49}
\end{align}
which satisfies the equality:

\begin{equation}
    E_{\text{FD}, 10} \approx \ESDTen
    \label{eq:50}
\end{equation}
and coincides with estimation~\eqn{eq:48}. This became possible because the thickness of the SD scintillator is 1.2~cm. In any other case estimation~\eqn{eq:49} would differ from~\eqn{eq:48} this or other way, depending on the thickness of a scintillator.

\subsection{The giant TA event}

It was reported that on May 27 2021 a giant event was registered at the TA~\cite{b:21}. Cosmic particle, dubbed Amaterasu, arrived at zenith angle $\theta = 38.6\degr \pm 0.4\degr$ and the resulting air shower was detected by SDs which recorded the value $S(800, \theta) = 530 \pm 57$. The energy of this unique event was estimated as $\ESD = (2.44 \pm 0.29) \times 10^{20}$~eV. Let us consider its calibration in the light of everything stated above. Square in \fign{f:16} represents the SD response $\log_{10}S(800, 38.6\degr) \approx 2.72 \pm 0.05$. According to calculations described in~\cite{b:14} (the ``$10^{19.5}$~eV'' gray band), the energy that corresponds to this signal is $E^{\text{TA}}_{\text{SD}} = 10^{2.72 - 1.86} \times 10^{19.5}$~eV $= 10^{20.36} \approx 2.29 \times 10^{20}$~eV. Within the errors of the experiment~\cite{b:21} and errors of transferring of the nomogram from \fign{f:6} to \fign{f:16}, this are, in fact, coinciding values. Our estimation results from data points at ``$10^{19.5}$~eV'' (downward triangles) in \fign{f:16}, and the resulting value is $\ESD = 10^{2.72 - 2.00} \times 10^{19.5} = 10^{20.22} \approx 1.75 \times 10^{20}$~eV. This difference in energy estimations~--- by factor $2.29/1.75 \approx 1.3$~--- arises from different zenith-angular dependencies of $S(800, \theta)$ shown in \fign{f:16}. It is difficult to say what has caused it. But the value of the difference, which is close to $\sec{}38.6\degr \approx 1.28$, draws attention. We have repeatedly noted this strange coincidence in both this work and in~\cite{b:4}. If one converts the $S(800, \theta)$ value to $S(800, 0\degr)$ using the two zenith-angular dependencies from \fign{f:16} denoted as ``$10^{19.5}$~eV'', then the resulting value will be $\ESD = 10^{2.80 - 2.07} \times 10^{19.5} = 10^{20.23} \approx 1.70 \times 10^{20}$~eV. The same estimation follows from the formula~\eqn{eq:33} with parameters listed in row~1 of Table~\ref{t:1} and the value of $\rho_s(800, 0\degr) = 10^{2.8}/3 \approx 210$~\pdens.

\section{Conclusion}

In this work the responses of the TA surface detectors were considered, calculated in air showers from primary protons with different energies and zenith angles of arrival using the \qgsii{} model. This model has proven itself during the investigation of the ``muon puzzle''~\cite{b:4, b:5}. It was demonstrated that estimation of primary energy $\ESD$ based on the readings of SD in vertical showers using the formula~\eqn{eq:1} does not contradict the calculations~\cite{b:14}. It leads to a better agreement between energy spectra from Yakutsk array and TA in terms of absolute value and form (\fign{f:1}). There is no such agreement in inclined showers. This may be due to incorrectly chosen unit of $\VEM = 2.05$~MeV in~\cite{b:14}, which does not reflect the real physical processes occurring in the scintillation detector during registration of EAS. Estimation of primary energy $\EFD$ obtained from the readings of FD, which is unconditionally prioritized at TA, is less than $\ESD$ by factor $\approx 1.264$. In the case of further confirmation of the arguments given in Section~4.2, this estimation may turn out to be incorrect. We will continue the cross-analysis of the calibrations of world EAS arrays in comparison with the Yakutsk array.

\section*{Funding}

This work was made within the framework of the state assignment No.\,122011800084-7 using the data obtained at The Unique Scientific Facility ``The D.\,D.~Krasilnikov Yakutsk Complex EAS Array'' (YEASA) (\url{https://ckp-rf.ru/catalog/usu/73611/}).

\section*{Conflict of interest}

The authors of this work declare that they have no conflict of interest.

\acknowledgements

Authors express their gratitude to the staff of the Separate structural unit YEASA of ShICRA SB RAS.

\bibliographystyle{apsmaik4-2}
\bibliography{ta_spectrum}

\begin{thebibliography}{25}%
\makeatletter
\providecommand \@ifxundefined [1]{%
 \@ifx{#1\undefined}
}%
\providecommand \@ifnum [1]{%
 \ifnum #1\expandafter \@firstoftwo
 \else \expandafter \@secondoftwo
 \fi
}%
\providecommand \@ifx [1]{%
 \ifx #1\expandafter \@firstoftwo
 \else \expandafter \@secondoftwo
 \fi
}%
\providecommand \natexlab [1]{#1}%
\providecommand \enquote  [1]{``#1''}%
\providecommand \bibnamefont  [1]{#1}%
\providecommand \bibfnamefont [1]{#1}%
\providecommand \citenamefont [1]{#1}%
\providecommand \href@noop [0]{\@secondoftwo}%
\providecommand \href [0]{\begingroup \@sanitize@url \@href}%
\providecommand \@href[1]{\@@startlink{#1}\@@href}%
\providecommand \@@href[1]{\endgroup#1\@@endlink}%
\providecommand \@sanitize@url [0]{\catcode `\\12\catcode `\$12\catcode
  `\&12\catcode `\#12\catcode `\^12\catcode `\_12\catcode `\%12\relax}%
\providecommand \@@startlink[1]{}%
\providecommand \@@endlink[0]{}%
\providecommand \url  [0]{\begingroup\@sanitize@url \@url }%
\providecommand \@url [1]{\endgroup\@href {#1}{\urlprefix }}%
\providecommand \urlprefix  [0]{URL }%
\providecommand \Eprint [0]{\href }%
\providecommand \doibase [0]{https://doi.org/}%
\providecommand \selectlanguage [0]{\@gobble}%
\providecommand \bibinfo  [0]{\@secondoftwo}%
\providecommand \bibfield  [0]{\@secondoftwo}%
\providecommand \translation [1]{[#1]}%
\providecommand \BibitemOpen [0]{}%
\providecommand \bibitemStop [0]{}%
\providecommand \bibitemNoStop [0]{.\EOS\space}%
\providecommand \EOS [0]{\spacefactor3000\relax}%
\providecommand \BibitemShut  [1]{\csname bibitem#1\endcsname}%
\let\auto@bib@innerbib\@empty
\bibitem [{\citenamefont {Coleman}\ \emph {et~al.}(2023)\citenamefont
  {Coleman}, \citenamefont {Eser}, \citenamefont {Mayotte}, \citenamefont
  {Sarazin}, \citenamefont {Schr\"{o}der}, \citenamefont {Soldin},
  \citenamefont {Venters}, \citenamefont {Aloisio}, \citenamefont
  {{Alvarez-Mu\~{n}iz}}, \citenamefont {{Alves Batista}}, \citenamefont
  {Bergman}, \citenamefont {Bertaina}, \citenamefont {Caccianiga},
  \citenamefont {Deligny}, \citenamefont {Dembinski}, \citenamefont {Denton}
  \emph {et~al.}}]{b:1}%
  \BibitemOpen
  \bibfield  {author} {\bibinfo {author} {\bibfnamefont {A.}~\bibnamefont
  {Coleman}}, \bibinfo {author} {\bibfnamefont {J.}~\bibnamefont {Eser}},
  \bibinfo {author} {\bibfnamefont {E.}~\bibnamefont {Mayotte}}, \bibinfo
  {author} {\bibfnamefont {F.}~\bibnamefont {Sarazin}}, \bibinfo {author}
  {\bibfnamefont {F.~G.}\ \bibnamefont {Schr\"{o}der}}, \bibinfo {author}
  {\bibfnamefont {D.}~\bibnamefont {Soldin}}, \bibinfo {author} {\bibfnamefont
  {T.~M.}\ \bibnamefont {Venters}}, \bibinfo {author} {\bibfnamefont
  {R.}~\bibnamefont {Aloisio}}, \bibinfo {author} {\bibfnamefont
  {J.}~\bibnamefont {{Alvarez-Mu\~{n}iz}}}, \bibinfo {author} {\bibfnamefont
  {R.}~\bibnamefont {{Alves Batista}}}, \bibinfo {author} {\bibfnamefont
  {D.}~\bibnamefont {Bergman}}, \bibinfo {author} {\bibfnamefont
  {M.}~\bibnamefont {Bertaina}}, \bibinfo {author} {\bibfnamefont
  {L.}~\bibnamefont {Caccianiga}}, \bibinfo {author} {\bibfnamefont
  {O.}~\bibnamefont {Deligny}}, \bibinfo {author} {\bibfnamefont {H.~P.}\
  \bibnamefont {Dembinski}}, \bibinfo {author} {\bibfnamefont {P.~B.}\
  \bibnamefont {Denton}}, \bibnamefont {et~al.},\ }\href
  {https://doi.org/10.1016/j.astropartphys.2023.102819} {\bibfield  {journal}
  {\bibinfo  {journal} {{Astroparticle Physics}}\ }\textbf {\bibinfo {volume}
  {149}},\ \bibinfo {pages} {102819} (\bibinfo {year} {2023})}\BibitemShut
  {NoStop}%
\bibitem [{\citenamefont {Glushkov}\ \emph {et~al.}(2018)\citenamefont
  {Glushkov}, \citenamefont {Pravdin},\ and\ \citenamefont {Saburov}}]{b:2}%
  \BibitemOpen
  \bibfield  {author} {\bibinfo {author} {\bibfnamefont {A.~V.}\ \bibnamefont
  {Glushkov}}, \bibinfo {author} {\bibfnamefont {M.~I.}\ \bibnamefont
  {Pravdin}},\ \bibnamefont {and}\ \bibinfo {author} {\bibfnamefont {A.~V.}\
  \bibnamefont {Saburov}},\ }\href {https://doi.org/10.1134/S106377881804004X}
  {\bibfield  {journal} {\bibinfo  {journal} {{Phys. At. Nucl.}}\ }\textbf
  {\bibinfo {volume} {81}},\ \bibinfo {pages} {575} (\bibinfo {year} {2018})},\
  \Eprint {https://arxiv.org/abs/2301.09654} {{arXiv}:2301.09654
  [{astro-ph.HE}]} \BibitemShut {NoStop}%
\bibitem [{\citenamefont {Glushkov}\ \emph {et~al.}(2024)\citenamefont
  {Glushkov}, \citenamefont {Saburov}, \citenamefont {Ksenofontov},\ and\
  \citenamefont {Lebedev}}]{b:4}%
  \BibitemOpen
  \bibfield  {author} {\bibinfo {author} {\bibfnamefont {A.~V.}\ \bibnamefont
  {Glushkov}}, \bibinfo {author} {\bibfnamefont {A.~V.}\ \bibnamefont
  {Saburov}}, \bibinfo {author} {\bibfnamefont {L.~T.}\ \bibnamefont
  {Ksenofontov}},\ \bibnamefont {and}\ \bibinfo {author} {\bibfnamefont
  {K.~G.}\ \bibnamefont {Lebedev}},\ }\href
  {https://doi.org/10.1134/S1063778824020121} {\bibfield  {journal} {\bibinfo
  {journal} {{Phys. At. Nucl.}}\ }\textbf {\bibinfo {volume} {87}},\ \bibinfo
  {pages} {25} (\bibinfo {year} {2024})},\ \Eprint
  {https://arxiv.org/abs/2306.17039} {{arXiv}:2306.17039 [{astro-ph.HE}]}
  \BibitemShut {NoStop}%
\bibitem [{\citenamefont {Glushkov}\ \emph {et~al.}(2023)\citenamefont
  {Glushkov}, \citenamefont {Saburov}, \citenamefont {Ksenofontov},\ and\
  \citenamefont {Lebedev}}]{b:5}%
  \BibitemOpen
  \bibfield  {author} {\bibinfo {author} {\bibfnamefont {A.~V.}\ \bibnamefont
  {Glushkov}}, \bibinfo {author} {\bibfnamefont {A.~V.}\ \bibnamefont
  {Saburov}}, \bibinfo {author} {\bibfnamefont {L.~T.}\ \bibnamefont
  {Ksenofontov}},\ \bibnamefont {and}\ \bibinfo {author} {\bibfnamefont
  {K.~G.}\ \bibnamefont {Lebedev}},\ }\href
  {https://doi.org/10.1134/S0021364023600726} {\bibfield  {journal} {\bibinfo
  {journal} {{JETP Lett.}}\ }\textbf {\bibinfo {volume} {117}},\ \bibinfo
  {pages} {645} (\bibinfo {year} {2023})},\ \Eprint
  {https://arxiv.org/abs/2304.13095} {{arXiv}:2304.13095 [{astro-ph.HE}]}
  \BibitemShut {NoStop}%
\bibitem [{\citenamefont {Arteaga-Vel\'{a}zquez}(2023)}]{b:6}%
  \BibitemOpen
  \bibfield  {author} {\bibinfo {author} {\bibfnamefont {J.~C.}\ \bibnamefont
  {Arteaga-Vel\'{a}zquez}},\ }in\ \href {https://doi.org/10.22323/1.444.0466}
  {\emph {\bibinfo {booktitle} {{Proceedings of the 38th International Cosmic
  Ray Conference}}}},\ Vol.\ \bibinfo {volume} {444},\ \bibinfo {editor} {Ed.
  by\ \bibinfo {editor} {\bibfnamefont {T.}~\bibnamefont {T.~Saito}}\
  \bibnamefont {and}\ \bibinfo {editor} {\bibfnamefont {K.}~\bibnamefont
  {Okumura}}}\ (\bibinfo  {publisher} {{SISSA}},\ \bibinfo {address} {{Nagoya,
  Japan}},\ \bibinfo {year} {2023})\ p.\ \bibinfo {pages} {466}\BibitemShut
  {NoStop}%
\bibitem [{\citenamefont {Nagano}\ \emph {et~al.}(1984)\citenamefont {Nagano},
  \citenamefont {Hara}, \citenamefont {Hatano}, \citenamefont {Hayashida},
  \citenamefont {Kawaguchi}, \citenamefont {Kamata}, \citenamefont {Kifune},\
  and\ \citenamefont {Mizumoto}}]{b:7}%
  \BibitemOpen
  \bibfield  {author} {\bibinfo {author} {\bibfnamefont {M.}~\bibnamefont
  {Nagano}}, \bibinfo {author} {\bibfnamefont {T.}~\bibnamefont {Hara}},
  \bibinfo {author} {\bibfnamefont {Y.}~\bibnamefont {Hatano}}, \bibinfo
  {author} {\bibfnamefont {N.}~\bibnamefont {Hayashida}}, \bibinfo {author}
  {\bibfnamefont {S.}~\bibnamefont {Kawaguchi}}, \bibinfo {author}
  {\bibfnamefont {K.}~\bibnamefont {Kamata}}, \bibinfo {author} {\bibfnamefont
  {K.}~\bibnamefont {Kifune}},\ \bibnamefont {and}\ \bibinfo {author}
  {\bibfnamefont {Y.}~\bibnamefont {Mizumoto}},\ }\href
  {https://doi.org/10.1088/0305-4616/10/9/016} {\bibfield  {journal} {\bibinfo
  {journal} {{J. Phys. G}}\ }\textbf {\bibinfo {volume} {10}},\ \bibinfo
  {pages} {1295} (\bibinfo {year} {1984})}\BibitemShut {NoStop}%
\bibitem [{\citenamefont {Nagano}\ \emph {et~al.}(1992)\citenamefont {Nagano},
  \citenamefont {Teshima}, \citenamefont {Matsubara}, \citenamefont {Dai},
  \citenamefont {Hara}, \citenamefont {Hayashida}, \citenamefont {Honda},
  \citenamefont {Ohoka},\ and\ \citenamefont {Yoshida}}]{b:8}%
  \BibitemOpen
  \bibfield  {author} {\bibinfo {author} {\bibfnamefont {M.}~\bibnamefont
  {Nagano}}, \bibinfo {author} {\bibfnamefont {M.}~\bibnamefont {Teshima}},
  \bibinfo {author} {\bibfnamefont {Y.}~\bibnamefont {Matsubara}}, \bibinfo
  {author} {\bibfnamefont {H.~Y.}\ \bibnamefont {Dai}}, \bibinfo {author}
  {\bibfnamefont {T.}~\bibnamefont {Hara}}, \bibinfo {author} {\bibfnamefont
  {N.}~\bibnamefont {Hayashida}}, \bibinfo {author} {\bibfnamefont
  {M.}~\bibnamefont {Honda}}, \bibinfo {author} {\bibfnamefont
  {H.}~\bibnamefont {Ohoka}},\ \bibnamefont {and}\ \bibinfo {author}
  {\bibfnamefont {S.}~\bibnamefont {Yoshida}},\ }\href
  {https://doi.org/10.1088/0954-3899/18/2/022} {\bibfield  {journal} {\bibinfo
  {journal} {{J. Phys. G}}\ }\textbf {\bibinfo {volume} {18}},\ \bibinfo
  {pages} {423} (\bibinfo {year} {1992})}\BibitemShut {NoStop}%
\bibitem [{\citenamefont {Shinozaki}(2006)}]{b:9}%
  \BibitemOpen
  \bibfield  {author} {\bibinfo {author} {\bibfnamefont {K.}~\bibnamefont
  {Shinozaki}} (\bibinfo {collaboration} {AGASA Collab.}),\ }\href
  {https://doi.org/10.1016/j.nuclphysbps.2005.07.002} {\bibfield  {journal}
  {\bibinfo  {journal} {{Nucl. Phys. B~--- Proc. Suppl.}}\ }\textbf {\bibinfo
  {volume} {151}},\ \bibinfo {pages} {3} (\bibinfo {year} {2006})}\BibitemShut
  {NoStop}%
\bibitem [{\citenamefont {Aartsen}\ \emph {et~al.}(2013)\citenamefont
  {Aartsen}, \citenamefont {Abbasi}, \citenamefont {Abdou}, \citenamefont
  {Ackermann}, \citenamefont {Adams}, \citenamefont {Aguilar}, \citenamefont
  {Ahlers}, \citenamefont {Altmann}, \citenamefont {Auffenberg}, \citenamefont
  {Bai}, \citenamefont {Baker}, \citenamefont {Barwick}, \citenamefont {Baum},
  \citenamefont {Bay}, \citenamefont {Beatty}, \citenamefont {Bechet} \emph
  {et~al.}}]{b:10}%
  \BibitemOpen
  \bibfield  {author} {\bibinfo {author} {\bibfnamefont {M.~G.}\ \bibnamefont
  {Aartsen}}, \bibinfo {author} {\bibfnamefont {R.}~\bibnamefont {Abbasi}},
  \bibinfo {author} {\bibfnamefont {Y.}~\bibnamefont {Abdou}}, \bibinfo
  {author} {\bibfnamefont {M.}~\bibnamefont {Ackermann}}, \bibinfo {author}
  {\bibfnamefont {J.}~\bibnamefont {Adams}}, \bibinfo {author} {\bibfnamefont
  {J.~A.}\ \bibnamefont {Aguilar}}, \bibinfo {author} {\bibfnamefont
  {M.}~\bibnamefont {Ahlers}}, \bibinfo {author} {\bibfnamefont
  {D.}~\bibnamefont {Altmann}}, \bibinfo {author} {\bibfnamefont
  {J.}~\bibnamefont {Auffenberg}}, \bibinfo {author} {\bibfnamefont
  {X.}~\bibnamefont {Bai}}, \bibinfo {author} {\bibfnamefont {M.}~\bibnamefont
  {Baker}}, \bibinfo {author} {\bibfnamefont {S.~W.}\ \bibnamefont {Barwick}},
  \bibinfo {author} {\bibfnamefont {V.}~\bibnamefont {Baum}}, \bibinfo {author}
  {\bibfnamefont {R.}~\bibnamefont {Bay}}, \bibinfo {author} {\bibfnamefont
  {J.~J.}\ \bibnamefont {Beatty}}, \bibinfo {author} {\bibfnamefont
  {S.}~\bibnamefont {Bechet}}, \bibnamefont {et~al.},\ }\href
  {https://doi.org/10.1103/PhysRevD.88.042004} {\bibfield  {journal} {\bibinfo
  {journal} {{Phys. Rev. D}}\ }\textbf {\bibinfo {volume} {86}},\ \bibinfo
  {pages} {042004} (\bibinfo {year} {2013})},\ \Eprint
  {https://arxiv.org/abs/1307.3795} {{arXiv}:1307.3795 [{astro-ph.HE}]}
  \BibitemShut {NoStop}%
\bibitem [{\citenamefont {Fenu}(2017)}]{b:11}%
  \BibitemOpen
  \bibfield  {author} {\bibinfo {author} {\bibfnamefont {F.}~\bibnamefont
  {Fenu}} (\bibinfo {collaboration} {{The Pierre Auger Collab.}}),\ }in\ \href
  {https://doi.org/10.22323/1.301.0486} {\emph {\bibinfo {booktitle}
  {{Proceedings of the 35th International Cosmic Ray Conference}}}},\ Vol.\
  \bibinfo {volume} {301},\ \bibinfo {editor} {Ed. by\ \bibinfo {editor}
  {\bibfnamefont {Y.-S.}\ \bibnamefont {Kwak}}, \bibinfo {editor}
  {\bibfnamefont {S.~H.}\ \bibnamefont {Lee}},\ \bibnamefont {and}\ \bibinfo
  {editor} {\bibfnamefont {S.}~\bibnamefont {Oh}}}\ (\bibinfo  {publisher}
  {SISSA},\ \bibinfo {address} {{Busan, Korea}},\ \bibinfo {year} {2017})\ p.\
  \bibinfo {pages} {486},\ \Eprint {https://arxiv.org/abs/1708.06592}
  {{arXiv}:1708.06592 [astro-ph.HE]} \BibitemShut {NoStop}%
\bibitem [{\citenamefont {Abu-Zayyad}\ \emph {et~al.}(2013)\citenamefont
  {Abu-Zayyad}, \citenamefont {Aida}, \citenamefont {Allen}, \citenamefont
  {Anderson}, \citenamefont {Azuma}, \citenamefont {Barcikowski}, \citenamefont
  {Belz}, \citenamefont {Bergman}, \citenamefont {Blake}, \citenamefont {Cady},
  \citenamefont {Cheon}, \citenamefont {Chiba}, \citenamefont {Chikawa},
  \citenamefont {Cho}, \citenamefont {Cho}, \citenamefont {Fujii} \emph
  {et~al.}}]{b:12}%
  \BibitemOpen
  \bibfield  {author} {\bibinfo {author} {\bibfnamefont {T.}~\bibnamefont
  {Abu-Zayyad}}, \bibinfo {author} {\bibfnamefont {R.}~\bibnamefont {Aida}},
  \bibinfo {author} {\bibfnamefont {M.}~\bibnamefont {Allen}}, \bibinfo
  {author} {\bibfnamefont {R.}~\bibnamefont {Anderson}}, \bibinfo {author}
  {\bibfnamefont {R.}~\bibnamefont {Azuma}}, \bibinfo {author} {\bibfnamefont
  {E.}~\bibnamefont {Barcikowski}}, \bibinfo {author} {\bibfnamefont {J.~W.}\
  \bibnamefont {Belz}}, \bibinfo {author} {\bibfnamefont {D.~R.}\ \bibnamefont
  {Bergman}}, \bibinfo {author} {\bibfnamefont {S.~A.}\ \bibnamefont {Blake}},
  \bibinfo {author} {\bibfnamefont {R.}~\bibnamefont {Cady}}, \bibinfo {author}
  {\bibfnamefont {B.~G.}\ \bibnamefont {Cheon}}, \bibinfo {author}
  {\bibfnamefont {J.}~\bibnamefont {Chiba}}, \bibinfo {author} {\bibfnamefont
  {M.}~\bibnamefont {Chikawa}}, \bibinfo {author} {\bibfnamefont {E.~J.}\
  \bibnamefont {Cho}}, \bibinfo {author} {\bibfnamefont {W.~R.}\ \bibnamefont
  {Cho}}, \bibinfo {author} {\bibfnamefont {H.}~\bibnamefont {Fujii}},
  \bibnamefont {et~al.},\ }\href {https://doi.org/10.1088/2041-8205/768/1/L1}
  {\bibfield  {journal} {\bibinfo  {journal} {{ApJ Lett.}}\ }\textbf {\bibinfo
  {volume} {768}},\ \bibinfo {pages} {L1} (\bibinfo {year} {2013})},\ \Eprint
  {https://arxiv.org/abs/1205.5067} {{arXiv}:1205.5067 [astro-ph.HE]}
  \BibitemShut {NoStop}%
\bibitem [{\citenamefont {Cunningham}\ \emph
  {et~al.}(1980{\natexlab{a}})\citenamefont {Cunningham}, \citenamefont
  {Lloyd-Evans}, \citenamefont {Pollock}, \citenamefont {Reid},\ and\
  \citenamefont {Watson}}]{b:13}%
  \BibitemOpen
  \bibfield  {author} {\bibinfo {author} {\bibfnamefont {G.}~\bibnamefont
  {Cunningham}}, \bibinfo {author} {\bibfnamefont {J.}~\bibnamefont
  {Lloyd-Evans}}, \bibinfo {author} {\bibfnamefont {A.~M.~T.}\ \bibnamefont
  {Pollock}}, \bibinfo {author} {\bibfnamefont {R.~J.}\ \bibnamefont {Reid}},\
  \bibnamefont {and}\ \bibinfo {author} {\bibfnamefont {A.~A.}\ \bibnamefont
  {Watson}},\ }\href {https://doi.org/10.1086/183201} {\bibfield  {journal}
  {\bibinfo  {journal} {{ApJ Lett.}}\ }\textbf {\bibinfo {volume} {236}},\
  \bibinfo {pages} {L71} (\bibinfo {year} {1980}{\natexlab{a}})}\BibitemShut
  {NoStop}%
\bibitem [{\citenamefont {Ivanov}(2012)}]{b:14}%
  \BibitemOpen
  \bibfield  {author} {\bibinfo {author} {\bibfnamefont {D.}~\bibnamefont
  {Ivanov}},\ }\emph {\bibinfo {title} {{Energy Spectrum Measured by the
  Telescope Array Surface Detector}}},\ \href
  {https://doi.org/10.7282/T3K35SG3} {Ph.D. thesis},\ \bibinfo  {school}
  {{Rutgers University}}, \bibinfo {address} {{New Brunswick NJ}} (\bibinfo
  {year} {2012})\BibitemShut {NoStop}%
\bibitem [{\citenamefont {Allison}\ \emph {et~al.}(2006)\citenamefont
  {Allison}, \citenamefont {Amako}, \citenamefont {Apostolakis}, \citenamefont
  {Araujo}, \citenamefont {{Arce Dubois}}, \citenamefont {Asai}, \citenamefont
  {Barrand}, \citenamefont {Capra}, \citenamefont {Chauvie}, \citenamefont
  {Chytracek}, \citenamefont {Cirrone}, \citenamefont {Cooperman},
  \citenamefont {Cosmo}, \citenamefont {Cuttone}, \citenamefont {Daquino},
  \citenamefont {Donszelmann} \emph {et~al.}}]{b:15}%
  \BibitemOpen
  \bibfield  {author} {\bibinfo {author} {\bibfnamefont {J.}~\bibnamefont
  {Allison}}, \bibinfo {author} {\bibfnamefont {K.}~\bibnamefont {Amako}},
  \bibinfo {author} {\bibfnamefont {J.}~\bibnamefont {Apostolakis}}, \bibinfo
  {author} {\bibfnamefont {H.}~\bibnamefont {Araujo}}, \bibinfo {author}
  {\bibfnamefont {P.}~\bibnamefont {{Arce Dubois}}}, \bibinfo {author}
  {\bibfnamefont {M.}~\bibnamefont {Asai}}, \bibinfo {author} {\bibfnamefont
  {G.}~\bibnamefont {Barrand}}, \bibinfo {author} {\bibfnamefont
  {R.}~\bibnamefont {Capra}}, \bibinfo {author} {\bibfnamefont
  {S.}~\bibnamefont {Chauvie}}, \bibinfo {author} {\bibfnamefont
  {R.}~\bibnamefont {Chytracek}}, \bibinfo {author} {\bibfnamefont {G.~A.~P.}\
  \bibnamefont {Cirrone}}, \bibinfo {author} {\bibfnamefont {G.}~\bibnamefont
  {Cooperman}}, \bibinfo {author} {\bibfnamefont {G.}~\bibnamefont {Cosmo}},
  \bibinfo {author} {\bibfnamefont {G.}~\bibnamefont {Cuttone}}, \bibinfo
  {author} {\bibfnamefont {G.~G.}\ \bibnamefont {Daquino}}, \bibinfo {author}
  {\bibfnamefont {M.}~\bibnamefont {Donszelmann}}, \bibnamefont {et~al.},\
  }\href {https://doi.org/10.1109/TNS.2006.869826} {\bibfield  {journal}
  {\bibinfo  {journal} {{IEEE Trans. Nucl. Sci.}}\ }\textbf {\bibinfo {volume}
  {53}},\ \bibinfo {pages} {270} (\bibinfo {year} {2006})}\BibitemShut
  {NoStop}%
\bibitem [{\citenamefont {Landau}(1944)}]{b:16}%
  \BibitemOpen
  \bibfield  {author} {\bibinfo {author} {\bibfnamefont {L.}~\bibnamefont
  {Landau}},\ }\href {https://doi.org/10.1016/B978-0-08-010586-4.50061-4}
  {\bibfield  {journal} {\bibinfo  {journal} {{J. Phys. USSR}}\ }\textbf
  {\bibinfo {volume} {8}},\ \bibinfo {pages} {201} (\bibinfo {year}
  {1944})}\BibitemShut {NoStop}%
\bibitem [{\citenamefont {Heck}\ \emph {et~al.}(1998)\citenamefont {Heck},
  \citenamefont {Knapp}, \citenamefont {Capdevielle}, \citenamefont {Schatz},\
  and\ \citenamefont {Thouw}}]{b:17}%
  \BibitemOpen
  \bibfield  {author} {\bibinfo {author} {\bibfnamefont {D.}~\bibnamefont
  {Heck}}, \bibinfo {author} {\bibfnamefont {J.}~\bibnamefont {Knapp}},
  \bibinfo {author} {\bibfnamefont {J.}~\bibnamefont {Capdevielle}}, \bibinfo
  {author} {\bibfnamefont {G.}~\bibnamefont {Schatz}},\ \bibnamefont {and}\
  \bibinfo {author} {\bibfnamefont {T.}~\bibnamefont {Thouw}},\ }\href
  {https://www.iap.kit.edu/corsika/70.php} {\emph {\bibinfo {title} {{CORSIKA:
  A Monte Carlo Code to Simulate Extensive Air Showers}}}},\ \bibinfo {type}
  {FZKA}\ \bibinfo {number} {6019}\ (\bibinfo  {institution}
  {{Forschungszentrum Karlsruhe}},\ \bibinfo {year} {1998})\BibitemShut
  {NoStop}%
\bibitem [{\citenamefont {Ostapchenko}(2011)}]{b:18}%
  \BibitemOpen
  \bibfield  {author} {\bibinfo {author} {\bibfnamefont {S.}~\bibnamefont
  {Ostapchenko}},\ }\href {https://doi.org/10.1103/PhysRevD.83.014018}
  {\bibfield  {journal} {\bibinfo  {journal} {{Phys. Rev. D}}\ }\textbf
  {\bibinfo {volume} {83}},\ \bibinfo {pages} {014018} (\bibinfo {year}
  {2011})},\ \Eprint {https://arxiv.org/abs/1010.1869} {{arXiv}:1010.1869
  [{hep-ph}]} \BibitemShut {NoStop}%
\bibitem [{\citenamefont {Glushkov}\ \emph {et~al.}(2014)\citenamefont
  {Glushkov}, \citenamefont {Pravdin},\ and\ \citenamefont {Saburov}}]{b:19}%
  \BibitemOpen
  \bibfield  {author} {\bibinfo {author} {\bibfnamefont {A.}~\bibnamefont
  {Glushkov}}, \bibinfo {author} {\bibfnamefont {M.~I.}\ \bibnamefont
  {Pravdin}},\ \bibnamefont {and}\ \bibinfo {author} {\bibfnamefont
  {A.}~\bibnamefont {Saburov}},\ }\href
  {https://doi.org/10.1134/S0021364014080086} {\bibfield  {journal} {\bibinfo
  {journal} {{JETP Lett.}}\ }\textbf {\bibinfo {volume} {99}},\ \bibinfo
  {pages} {431} (\bibinfo {year} {2014})}\BibitemShut {NoStop}%
\bibitem [{\citenamefont {Ferrari}\ \emph {et~al.}(2005)\citenamefont
  {Ferrari}, \citenamefont {Sala}, \citenamefont {Paola}, \citenamefont
  {Fass\`{o}},\ and\ \citenamefont {Ranft}}]{b:FLUKA}%
  \BibitemOpen
  \bibfield  {author} {\bibinfo {author} {\bibfnamefont {A.}~\bibnamefont
  {Ferrari}}, \bibinfo {author} {\bibfnamefont {P.}~\bibnamefont {Sala}},
  \bibinfo {author} {\bibfnamefont {R.}~\bibnamefont {Paola}}, \bibinfo
  {author} {\bibfnamefont {A.}~\bibnamefont {Fass\`{o}}},\ \bibnamefont {and}\
  \bibinfo {author} {\bibfnamefont {J.}~\bibnamefont {Ranft}},\ }\href
  {https://doi.org/10.5170/CERN-2005-010} {\emph {\bibinfo {title} {{FLUKA: A
  multi-particle transport code (program version 2005)}}}},\ \bibinfo {type}
  {Report}\ \bibinfo {number} {CERN-2005-010}\ (\bibinfo  {institution}
  {{CERN}},\ \bibinfo {year} {2005})\BibitemShut {NoStop}%
\bibitem [{\citenamefont {{Saburov, A.}}(2018)}]{b:20}%
  \BibitemOpen
  \bibfield  {author} {\bibinfo {author} {\bibnamefont {{Saburov, A.}}},\
  }\emph {\bibinfo {title} {{Lateral distribution of particles in EASs with
  energy above $10^{17}$~eV according to the Yakutsk array data}}},\ \href@noop
  {} {Ph.D. thesis},\ \bibinfo  {school} {{INR RAS}}, \bibinfo {address}
  {{Moscow}} (\bibinfo {year} {2018}),\ \bibinfo {note} {{[in
  Russian]}}\BibitemShut {NoStop}%
\bibitem [{\citenamefont {Abbasi}\ \emph {et~al.}(2023)\citenamefont {Abbasi},
  \citenamefont {Allen}, \citenamefont {Arimura}, \citenamefont {Belz},
  \citenamefont {Bergman}, \citenamefont {Blake}, \citenamefont {Shin},
  \citenamefont {Buckland}, \citenamefont {Cheon}, \citenamefont {Fujii},
  \citenamefont {Fujisue}, \citenamefont {Fujita}, \citenamefont {Fukushima},
  \citenamefont {Furlich}, \citenamefont {Gerber}, \citenamefont {Globus} \emph
  {et~al.}}]{b:21}%
  \BibitemOpen
  \bibfield  {author} {\bibinfo {author} {\bibfnamefont {R.~U.}\ \bibnamefont
  {Abbasi}}, \bibinfo {author} {\bibfnamefont {M.~G.}\ \bibnamefont {Allen}},
  \bibinfo {author} {\bibfnamefont {R.}~\bibnamefont {Arimura}}, \bibinfo
  {author} {\bibfnamefont {J.~W.}\ \bibnamefont {Belz}}, \bibinfo {author}
  {\bibfnamefont {D.~R.}\ \bibnamefont {Bergman}}, \bibinfo {author}
  {\bibfnamefont {S.~A.}\ \bibnamefont {Blake}}, \bibinfo {author}
  {\bibfnamefont {B.~K.}\ \bibnamefont {Shin}}, \bibinfo {author}
  {\bibfnamefont {I.~J.}\ \bibnamefont {Buckland}}, \bibinfo {author}
  {\bibfnamefont {B.~G.}\ \bibnamefont {Cheon}}, \bibinfo {author}
  {\bibfnamefont {T.}~\bibnamefont {Fujii}}, \bibinfo {author} {\bibfnamefont
  {K.}~\bibnamefont {Fujisue}}, \bibinfo {author} {\bibfnamefont
  {K.}~\bibnamefont {Fujita}}, \bibinfo {author} {\bibfnamefont
  {M.}~\bibnamefont {Fukushima}}, \bibinfo {author} {\bibfnamefont {G.~D.}\
  \bibnamefont {Furlich}}, \bibinfo {author} {\bibfnamefont {Z.~R.}\
  \bibnamefont {Gerber}}, \bibinfo {author} {\bibfnamefont {N.}~\bibnamefont
  {Globus}}, \bibnamefont {et~al.},\ }\href
  {https://doi.org/10.1126/science.abo5095} {\bibfield  {journal} {\bibinfo
  {journal} {{Science}}\ }\textbf {\bibinfo {volume} {382}},\ \bibinfo {pages}
  {903} (\bibinfo {year} {2023})},\ \Eprint {https://arxiv.org/abs/2311.14231}
  {{arXiv}:2311.14231 [{astro-ph.HE}]} \BibitemShut {NoStop}%
\bibitem [{\citenamefont {Tennent}(1967)}]{b:22}%
  \BibitemOpen
  \bibfield  {author} {\bibinfo {author} {\bibfnamefont {R.~M.}\ \bibnamefont
  {Tennent}},\ }\href {https://doi.org/10.1088/0370-1328/92/3/315} {\bibfield
  {journal} {\bibinfo  {journal} {Proc. of Phys. Soc.}\ }\textbf {\bibinfo
  {volume} {92}},\ \bibinfo {pages} {622} (\bibinfo {year} {1967})}\BibitemShut
  {NoStop}%
\bibitem [{\citenamefont {Cunningham}\ \emph
  {et~al.}(1980{\natexlab{b}})\citenamefont {Cunningham}, \citenamefont {Edge},
  \citenamefont {England}, \citenamefont {Evans}, \citenamefont {Hollows},
  \citenamefont {Hopper}, \citenamefont {Liversedge}, \citenamefont
  {Lloyd-Evans}, \citenamefont {Ogden}, \citenamefont {Patel}, \citenamefont
  {Pearce}, \citenamefont {Pollock}, \citenamefont {Reid}, \citenamefont
  {Tennent}, \citenamefont {Walker}, \citenamefont {Watson}, \citenamefont
  {Wilson},\ and\ \citenamefont {Wray}}]{b:23}%
  \BibitemOpen
  \bibfield  {author} {\bibinfo {author} {\bibfnamefont {G.}~\bibnamefont
  {Cunningham}}, \bibinfo {author} {\bibfnamefont {D.~M.}\ \bibnamefont
  {Edge}}, \bibinfo {author} {\bibfnamefont {D.}~\bibnamefont {England}},
  \bibinfo {author} {\bibfnamefont {A.~C.}\ \bibnamefont {Evans}}, \bibinfo
  {author} {\bibfnamefont {J.~D.}\ \bibnamefont {Hollows}}, \bibinfo {author}
  {\bibfnamefont {J.}~\bibnamefont {Hopper}, \bibfnamefont {S.~J.~Lapikens}},
  \bibinfo {author} {\bibfnamefont {B.}~\bibnamefont {Liversedge}}, \bibinfo
  {author} {\bibfnamefont {J.}~\bibnamefont {Lloyd-Evans}}, \bibinfo {author}
  {\bibfnamefont {P.}~\bibnamefont {Ogden}}, \bibinfo {author} {\bibfnamefont
  {M.}~\bibnamefont {Patel}}, \bibinfo {author} {\bibfnamefont
  {D.}~\bibnamefont {Pearce}}, \bibinfo {author} {\bibfnamefont {A.~M.~T.}\
  \bibnamefont {Pollock}}, \bibinfo {author} {\bibfnamefont {R.~J.~O.}\
  \bibnamefont {Reid}}, \bibinfo {author} {\bibfnamefont {R.~M.}\ \bibnamefont
  {Tennent}}, \bibinfo {author} {\bibfnamefont {R.}~\bibnamefont {Walker}},
  \bibinfo {author} {\bibfnamefont {A.~A.}\ \bibnamefont {Watson}},
  \bibnamefont {et~al.},\ }in\ \href
  {https://inspirehep.net/literature/1293873} {\emph {\bibinfo {booktitle}
  {{Catalogue of Highest Energy Cosmic Rays: Giant Extensive Air Showers. No.
  1: Volcano Ranch and Haverah Park}}}},\ \bibinfo {series and number}
  {\bibinfo {number} {CHECR-1-1980}},\ \bibinfo {editor} {Ed. by\ \bibinfo
  {editor} {\bibfnamefont {M.}~\bibnamefont {Wada}}}\ (\bibinfo  {publisher}
  {{World Data Center C2 for Cosmic Rays The Institute of Physical and Chemical
  Research}},\ \bibinfo {address} {Itabashi, Tokyo Japan},\ \bibinfo {year}
  {1980})\ p.~\bibinfo {pages} {61}\BibitemShut {NoStop}%
\bibitem [{\citenamefont {Aab}\ \emph {et~al.}(2015)\citenamefont {Aab},
  \citenamefont {Abreu}, \citenamefont {Aglietta}, \citenamefont {Ahn},
  \citenamefont {Al~Samarai}, \citenamefont {Albert}, \citenamefont
  {Albuquerqu}, \citenamefont {Allekotte}, \citenamefont {Allen}, \citenamefont
  {Allison}, \citenamefont {Almela}, \citenamefont {Alvarez~Castillo},
  \citenamefont {Alvarez-Mu\~{n}iz}, \citenamefont {Alves~Batista},
  \citenamefont {Ambrosio}, \citenamefont {Aminaei} \emph {et~al.}}]{b:24}%
  \BibitemOpen
  \bibfield  {author} {\bibinfo {author} {\bibfnamefont {A.}~\bibnamefont
  {Aab}}, \bibinfo {author} {\bibfnamefont {P.}~\bibnamefont {Abreu}}, \bibinfo
  {author} {\bibfnamefont {M.}~\bibnamefont {Aglietta}}, \bibinfo {author}
  {\bibfnamefont {E.~J.}\ \bibnamefont {Ahn}}, \bibinfo {author} {\bibfnamefont
  {I.}~\bibnamefont {Al~Samarai}}, \bibinfo {author} {\bibfnamefont {J.~N.}\
  \bibnamefont {Albert}}, \bibinfo {author} {\bibfnamefont {I.~F.~M.}\
  \bibnamefont {Albuquerqu}}, \bibinfo {author} {\bibfnamefont
  {I.}~\bibnamefont {Allekotte}}, \bibinfo {author} {\bibfnamefont
  {J.}~\bibnamefont {Allen}}, \bibinfo {author} {\bibfnamefont
  {P.}~\bibnamefont {Allison}}, \bibinfo {author} {\bibfnamefont
  {A.}~\bibnamefont {Almela}}, \bibinfo {author} {\bibfnamefont
  {J.}~\bibnamefont {Alvarez~Castillo}}, \bibinfo {author} {\bibfnamefont
  {J.}~\bibnamefont {Alvarez-Mu\~{n}iz}}, \bibinfo {author} {\bibfnamefont
  {R.}~\bibnamefont {Alves~Batista}}, \bibinfo {author} {\bibfnamefont
  {M.}~\bibnamefont {Ambrosio}}, \bibinfo {author} {\bibfnamefont
  {A.}~\bibnamefont {Aminaei}}, \bibnamefont {et~al.},\ }\href
  {https://doi.org/10.1016/j.nima.2015.06.058} {\bibfield  {journal} {\bibinfo
  {journal} {{Nucl. Instr. Methods A}}\ }\textbf {\bibinfo {volume} {798}},\
  \bibinfo {pages} {172} (\bibinfo {year} {2015})},\ \Eprint
  {https://arxiv.org/abs/1502.01323} {{arXiv}:1502.01323 [{astro-ph.IM}]}
  \BibitemShut {NoStop}%
\bibitem [{\citenamefont {Gaisser}\ and\ \citenamefont {Hillas}(1977)}]{b:25}%
  \BibitemOpen
  \bibfield  {author} {\bibinfo {author} {\bibfnamefont {T.~K.}\ \bibnamefont
  {Gaisser}}\ \bibnamefont {and}\ \bibinfo {author} {\bibfnamefont {A.~M.}\
  \bibnamefont {Hillas}},\ }in\ \href
  {https://ui.adsabs.harvard.edu/abs/1977ICRC....8..353G/abstract} {\emph
  {\bibinfo {booktitle} {{Proceedings of the 15th International Cosmic Ray
  Conference}}}},\ Vol.~\bibinfo {volume} {8}\ (\bibinfo  {publisher}
  {{UIPAP}},\ \bibinfo {address} {{Plovdiv, Bulgaria}},\ \bibinfo {year}
  {1977})\ p.\ \bibinfo {pages} {353}\BibitemShut {NoStop}%
\end{thebibliography}%

\end{document}